\documentclass[%
 reprint,
 amsmath,amssymb,
 aps,
]{revtex4-2}

\usepackage{graphicx}% Include figure files
\usepackage[svgnames,dvipsnames,x11names,table]{xcolor}
\usepackage{dcolumn}% Align table columns on decimal point
\usepackage{bm}% bold math
\usepackage{hyperref}% add hypertext capabilities
\usepackage[retainorgcmds]{IEEEtrantools}
\usepackage[neverdecrease]{paralist}
\usepackage{subfigure}
\usepackage{float}

\newcommand{\ud}{\mathrm{d}}

\begin{document}

%\preprint{APS/123-QED}

\title{Constraining fuzzy dark matter with the 21-cm power spectrum\\ from Cosmic Dawn and Reionization}% Force line breaks with \\

\author{Shihang Liu}
 \email{sehighs@163.com}
%\affiliation{School of Physical Science and Technology, Guangxi University, Nanning 530004, China}
\author{Yilin Liu}%
 %\email{one.liu01@outlook.com}
\author{Bowen Peng}
\author{Mengzhou Xie}
\author{Zelong Liu}
\author{Bohua Li}
 \email{bohuali@gxu.edu.cn}
\affiliation{Guangxi Key Laboratory for Relativistic Astrophysics, 
School of Physical Science and Technology, Guangxi University, \\
Nanning 530004, People’s Republic of China}

\author{Yi Mao}
\affiliation{Department of Astronomy, Tsinghua University, Beijing 100084, 
People’s Republic of China}

\date{\today}% It is always \today, today,

\begin{abstract}

The 21-cm signals from Cosmic Dawn and the Epoch of Reionization
contain valuable information on cosmological structure formation
dominated by dark matter.
Measurements of the 21-cm power spectrum
can thus probe certain dark matter candidates.
Here we investigate the impacts of fuzzy dark matter (FDM) on the 21-cm signals,
taking into account both the linear matter power spectrum
and the halo mass function (HMF) in FDM cosmologies.
The full FDM dynamics are implemented in reionization simulations,
along with a new ansatz on modulation of the FDM HMF by the linear overdensity.
Not only does the suppression of FDM halos on small scales
give rise to delay of the signature epochs during cosmic reionization,
but these epochs are also shortened relative to the cold dark matter cosmology.
In addition, we find that while the FDM effects on the 21-cm power spectrum
are dominated by its linear dynamics early in Cosmic Dawn,
a correct FDM HMF resulting from nonlinear wave dynamics
must be considered when X-ray heating begins.
We forecast the constraints on the FDM model parameters
from upcoming 21-cm power spectrum measurements by SKA1-Low (central area).
In FDM cosmologies with $m_\mathrm{FDM}=10^{-21}$\,eV,
SKA1-Low will be able to constrain the boson mass
to within $\sim10$\% at 2$\sigma$ confidence with a mock 1080-hour observation,
if the ionizing efficiency is mass independent.
However, our results show that realistic astrophysical processes
are degenerate with the FDM effects,
which shall severely loosen the constraints on the boson mass
from 21-cm power spectrum data alone.

%In addition, the HERA and SKA1-Low central area can also have good constraints on the underlying parameter of the FDM universe: the mass function index, although this parameter has little effect on the history of reionization.

\end{abstract}

\maketitle

%\tableofcontents

\section{Introduction}

%Dark matter has been discovered for decades, and the cold dark matter (CDM) is widely accepted as a component of the standard cosmological model. Although cosmic microwave background (CMB) measurement have confirmed that the energy density of CDM is about 1/5 of the total energy density of the universe \cite{aghanim2020planck}, and we have modeled that dark matter are almost affected by gravitation, the intrinsic nature of dark matter particles has remain unknown.

The physical nature of cosmological dark matter is still unknown.
In the meantime, cold dark matter (CDM), as the longtime standard dark matter model,
has faced serious challenges over the past two decades,
arising from the increasing evidence of discrepancies
between its theoretical predictions and observational data on small scales
\cite[e.g.,][]{bullock2017small,weinberg2015cold}.
%Well-known examples include
%The standard cosmological model including CDM has achieved great succes, but has outstanding problems interpreting the small scale structure of the universe, which can be summarized as 
%(i) the missing satellites problem, (ii) the cusp-core problem
%and (iii) the ``too big to fail'' problem, etc \cite{bullock2017small,weinberg2015cold}.
%The missing satellite problem is that galaxies like the Milky Way should have more satellite galaxies than we can observe, and we can solve this problem by hypothesize these satellite dwarf galaxies have lost stars due to tidal action, making them impossible to observe. But the solution creates a tension by the ‘too big to fail’ problem. $\Lambda$CDM model predicted too giant mass for these satellite galaxies to be lost stars. As for cusp-core problem, it means the halo profile with $\Lambda$CDM case predicting a very high density peak within galaxies core, but the core density should be flat enough.
While the validity of these challenges is still under debate, 
they have motivated various alternative dark matter candidates.

Fuzzy dark matter (FDM) or scalar field dark matter \cite[e.g.,][]{peebles2000fluid,goodman2000repulsive,hu2000fuzzy,boyle2002spintessence,2010PhRvD..81l3530A,li2014cosmological,2016PhR...643....1M,hui2017ultralight},
currently one of the most promising alternatives,
consists of ultralight bosons with masses $\sim 10^{-22}\,\text{eV}$. 
As a result, the de Broglie wavelength of the FDM particles
can reach $\sim$\,kpc scales. 
The clustering properties of FDM are mostly identical
with those of CDM on larger scales,
but reveal a distinct wave nature of FDM on scales
below its de Broglie wavelength.
FDM is therefore also called wave dark matter in the literature
\cite{schive2014cosmic,schive2014understanding,hui2021wave}.
In FDM cosmologies, the growth of cosmic structure
is suppressed below its de Broglie wavelength, providing potential solutions
to the aforementioned small-scale CDM problems
\cite{hu2000fuzzy,hlozek2015search, 2025SCPMA..6880409Y}.
Physically, suppression of small-scale structures
occurs in both linear and nonlinear stages of structure formation. 
In the linear regime, the power spectrum of the FDM density fluctuations
is truncated below its effective Jeans scale \cite{khlopov1985gravitational,hu2000fuzzy}. 
%affecting the dark matter halo formation at later times.
In the nonlinear regime, a ``quantum pressure'' term appears
in the momentum equation of FDM \cite{2011PhRvD..84d3531C,2012MNRAS.422..135R}.
%that governs the full dynamics 
This pressure term counteracts gravitational collapse, 
thereby quenching the formation of dark matter halos on the smallest scales.
Altogether, the abundance of low-mass halos decreases
relative to that in the CDM case, 
resulting in a modified halo mass function (HMF)
\cite{schive2016contrasting, may2021structure,may2023halo}.
This brings forth a series of testable effects
on the formation of the first luminous objects, 
which can be probed by high-redshift observations
of cosmic dawn (CD) and the epoch of reionization (EoR)
\cite{hui2017ultralight,lidz2018implications, 2023NatAs...7..731B}. 

The 21-cm line of neutral hydrogen (H\,{\small I})
in the intergalactic medium (IGM) is considered to be
one of the most promising probes of the Dark Ages and CD/EoR
\cite[e.g.,][]{furlanetto2006cosmology,2008PhRvD..78b3529M,pritchard201221,barkana2016rise}.
The brightness temperature of the redshifted 21-cm radiation maps
the H\,{\small I} gas in the high-$z$ IGM tomographically,
encoding crucial information on the timing and morphology of cosmic reionization.
The 21-cm brightness temperature signal is often presented in terms of
its global (sky-averaged) signal
and its spatial fluctuations described by the power spectrum.

The impact of FDM on the 21-cm power spectrum has previously been studied in
\cite{lidz2018implications,nebrin2019fuzzy,jones2021fuzzy,sarkar2022exploring,flitter2022closing,2022PhRvD.105h3011G}. 
Reference~\cite{lidz2018implications} found that
if the spin temperature of the 21-cm transition is required to
be coupled to the kinetic temperature of the H\,{\small I} gas by $z = 20$,
then a lower limit must be placed on the mass of the ultralight boson,
$m_\mathrm{FDM} \geq 5 \times 10^{-21}$\,eV.
Reference~\cite{jones2021fuzzy} used the Fisher matrix formalism
to forecast the capability of the Hydrogen Epoch of Reionization Array (HERA) \cite{deboer2017hydrogen,dillon2016redundant}
to constrain $m_\mathrm{FDM}$.
They found that in an FDM universe with $m_\mathrm{FDM} = 10^{-21}\,\mathrm{eV}$, 
HERA can determine the boson mass to within 20$\%$ at 2$\sigma$ confidence.
Reference~\cite{sarkar2022exploring} performed a similar analysis
and found that HERA can probe FDM
of masses up to $m_\mathrm{FDM}\sim 10^{-19}-10^{-18}\,\mathrm{eV}$.
Reference~\cite{flitter2022closing} showed that
for the mass window of $10^{-25}-10^{-23}\,\mathrm{eV}$,
HERA will be sensitive to FDM fractions as low as 1\%.
Reference~\cite{2022PhRvD.105h3011G} generated mock observations
for the Square Kilometre Array (SKA) \cite{2013ExA....36..235M}
based on a $\Lambda$CDM cosmology to forecast its constraints on $m_\mathrm{FDM}$.
In all these works, the FDM effects on structure formation
are either modeled on the level of linear dynamics only
(using the FDM power spectrum \cite{jones2021fuzzy}),
or modeled on the nonlinear level
by a halo mass function derived from the excursion-set formalism
\cite{marsh2014model,lidz2018implications,nebrin2019fuzzy,2022PhRvD.105h3011G}.
However, such HMFs fail to match the results
from cosmological simulations of the full FDM dynamics (linear+nonlinear)
\cite{may2021structure,may2023halo}, which indicates that
the excursion-set formalism is not suitable for FDM cosmologies.
Therefore, a self-consistent prescription for the FDM HMF
must result from full numerical simulations,
even if the statistical significance of the latter is still limited currently
\cite{may2023halo}.

In this paper, we consider the full FDM dynamics
and their impacts on the 21-cm signals from CD and the EoR,
taking into account both the linear FDM power spectrum
and the FDM HMF from nonlinear dynamics.
We modify the publicly available \verb|21cmFAST| code
\cite{mesinger201121cmfast,greig2017simultaneously,park2019inferring,munoz2022impact}
to simulate CD/EoR and calculate the 21-cm brightness temperature. 
We characterize the differences
between the 21-cm signals in FDM cosmologies and those in a CDM cosmology,
and examine the effects of varying $m_\mathrm{FDM}$ and other key parameters.
We forecast the prospects for detecting or constraining FDM
with the low-frequency array of SKA (SKA1-Low) or HERA,
based on the Fisher matrix formalism \cite{tegmark1997karhunen}.
Our results here complement existing observational constraints on the FDM model,
e.g., those from the Lyman-alpha (Ly$\alpha$) forest
\cite{2017PhRvL.119c1302I,2021PhRvL.126g1302R}
and the dark matter halo core in M87 \cite{2022ApJ...929...17D}.

The rest of this paper is organized as follows. 
In \S\ref{sec:model}, we introduce how we model the 21-cm signals in FDM cosmologies
as well as our simulations of CD/EoR.
We show the simulation results in \S\ref{sec:results}.
Based on the resulting 21-cm power spectra from CD/EoR in FDM cosmologies,
we present a Fisher matrix forecast on measuring the FDM parameters in \S\ref{sec:forecast}.
In \S\ref{sec:discussion}, we extend the astrophysical sector of the CD/EoR model
to show the degeneracy between the effects of FDM dynamics and the astrophysical effects,
and we discuss how this degeneracy impacts the constraints on FDM parameters.
We conclude in \S\ref{sec:conclusion}.

\section{\label{sec:model}MODEL and simulation}

We calculate the 21-cm brightness temperature signal from CD/EoR
using seminumerical simulations.
The signal is defined as the following differential brightness temperature
at frequency $\nu$ \cite[e.g.,][]{madau199721,furlanetto2006cosmology}:
\begin{equation}\label{eq:delta_Tb}
    \begin{split}
    \delta T_\mathrm{b}(\nu) & \equiv \frac{T_\mathrm{S} - T_\gamma}{1 + z} (1 - e^{-\tau_\nu})\\
& \approx 27\,x_\mathrm{H\,{I}}\,(1 + \delta_\mathrm{b})
\left( \frac{H}{\mathrm{d}v_r/\mathrm{d}r + H} \right) \left( 1 - \frac{T_\gamma}{T_\mathrm{S}} \right)\\
&\times \left( \frac{1 + z}{10} \frac{0.15}{\Omega_\mathrm{M} h^2} \right)^{1/2} 
\left( \frac{\Omega_\mathrm{b} h^2}{0.023} \right) ~\mathrm{mK}.
    \end{split}
\end{equation}
Here $T_\gamma$ is the CMB temperature
along the line of sight (LoS) \footnote{Here we do not consider
an excess radio background in modeling the 21-cm brightness temperature
in Eq.~(\ref{eq:delta_Tb}). We save this for future work.},
$T_\mathrm{S}$ is the spin temperature of the emitting cloud
at redshift $z = \nu_{21} /\nu-1$ ($\nu_{21}\approx 1420$\,MHz),
$\tau_\nu$ is its 21-cm optical depth, and $\ud v_r /\ud r$ is the comoving gradient
of the LoS component of its comoving velocity.
In this paper, we only consider the linear redshift-space distortion effect, 
assuming $\ud v_r /\ud r \ll H$,
which is generally true for the pertinent redshifts and scales
\cite{mao2012redshift,2021ApJ...918...14L}.
As usual, $\delta_\mathrm{b}(\vec x, z) \equiv \rho_\mathrm{b}/\bar{\rho}_\mathrm{b}-1$
is the (Eulerian) baryon overdensity and $H(z)$ is the Hubble parameter.

We modify the publicly available \verb|21cmFAST| code
to simulate cosmic dawn and the epoch of reionization
\cite{mesinger201121cmfast,murray202021cmfast}.
%The code includes an approximate treatment of Ly-$\alpha$ background photons, 
In addition to the ionization history of the IGM,
we compute the full evolution of the H\,{\small I} spin temperature
beginning at $z=35$.
%responsible for coupling the spin temperature of the 21 cm line to the gas temperature, 
Contributions from early x-ray sources to ionization and heating of the IGM
are also taken into account in our seminumerical simulations.

We introduce our model of FDM structure formation in \S\ref{sec:FDM}, 
which addresses the initial conditions and the halo mass function. 
\S\ref{sec:Astro} provides a brief overview of the astrophysical model 
used in \verb|21cmFAST|, describing the key astrophysical parameters.
In \S\ref{ssec:simulation}, we summarize the suite of \verb|21cmFAST| simulations
performed in this work.

\subsection{\label{sec:FDM}Structure formation with FDM}

Here we describe our modifications to the \verb|21cmFAST| implementations
of structure formation, taking into account the full FDM dynamics. 
In our model, the linear matter power spectrum of FDM is suppressed on small scales, 
serving as the initial conditions for our seminumerical simulations
(\S\ref{sssec:ICs}).
The nonlinear effects of FDM on halo formation are modeled by
(i) a fitting formula for the halo mass function, 
adopted from full FDM numerical simulations
\cite{may2021structure, may2023halo},
(ii) an ansatz that treats the density-modulated environmental effects
in FDM cosmologies, which we introduce for the first time
(\S\ref{sssec:FDMHMF}).

\subsubsection{Initial Conditions}\label{sssec:ICs}

\begin{figure}[htp]
\includegraphics[width = 8.5cm]{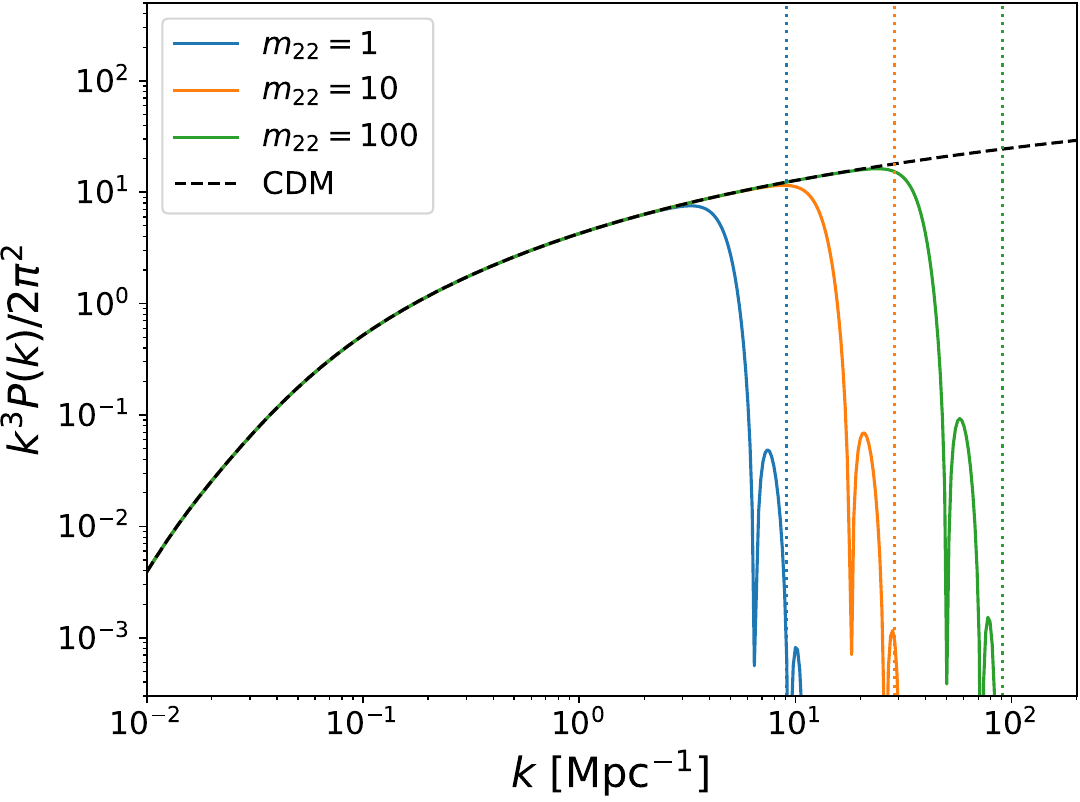}% Here is how to import EPS art
\caption{Linear matter power spectra extrapolated to $z=0$ in FDM and CDM cosmologies. 
The FDM power spectra are obtained using Eq.~(\ref{eq:ps}) 
for various boson masses, $m_{22}=1$, 10 and 100. 
The vertical dotted lines indicate
the FDM Jeans wave number at matter-radiation equality, $k_\mathrm{J,\,eq}$.}
\label{fig:ICs}
\end{figure}

We adopt the fitting formula for the linear FDM power spectrum
in Ref.~\cite{hu2000fuzzy}, expressed as
\begin{equation}
    \frac{P_{\mathrm{FDM}}(k,z)}{P_\mathrm{CDM}(k,z)} 
    = \left[ \frac{\cos(x^3(k))}{1 + x^8(k)} \right]^2.
    \label{eq:ps}
\end{equation}
Here $P_\mathrm{CDM}(k,z)$ is the CDM power spectrum
and $x(k)\equiv 1.61\,m_{22}^{1/18}\,(k/k_\mathrm{J,\,eq})$, 
where $m_{22}\equiv m_{\mathrm{FDM}}/(10^{-22}\,\mathrm{eV})$
is the normalized FDM particle mass
and $k_\mathrm{J,\,eq} = 9\,m_{22}^{1/2}\,\mathrm{Mpc}^{-1}$
is the FDM Jeans wave number at matter-radiation equality
\cite{hu2000fuzzy, khlopov1985gravitational}.
In Fig.~\ref{fig:ICs}, we show the present-day FDM linear power spectra
for various $m_{22}$.

%Ref.~\cite{hu2000fuzzy} also defines another characteristic wavenumber concerning the cutoff effect: $k_{1/2}$, the scale at which the power drops by a factor of two.
%It turns out that $k_{1/2}\sim 4.5\,m_{22}^{4/9}~\mathrm{Mpc}^{-1}\sim 0.5\,k_\mathrm{J,\,eq}$. 
%The corresponding mass scale is then given by $M_{1/2}\equiv(4\pi \bar{\rho}_\mathrm{m}/3) \left( \pi/k_{1/2} \right)^3 = 5.6 \times 10^{10}\,m_{22}^{-4/3}\,M_{\odot}$, where $\bar{\rho}_\mathrm{m}$ is the mean comoving matter density.

\subsubsection{The FDM halo mass function}\label{sssec:FDMHMF}

In addition to the small-scale cutoff in the initial conditions,
the nonlinear dynamical effects of FDM due to its wave nature
can alter the shape and the redshift evolution of the halo mass function
relative to the CDM case.
The results of the numerical simulations in Refs.~\cite{may2021structure,may2023halo}
show that the FDM HMF can be approximated by the following fitting formula,
first proposed in Ref.~\cite{schive2016contrasting}:
\begin{equation}\label{eq:HMF}
    \frac{\ud n}{\ud m} \bigg|_\mathrm{FDM} (m, z)
    = \frac{\ud n}{\ud m} \bigg|_\mathrm{CDM} (m, z) 
    \left[1 + \left( \frac{m}{M_0} \right)^{\alpha}\,\right]^{-2.2},
\end{equation}
where $M_0\equiv 1.6 \times 10^{10}\,m_{22}^{-4/3}\,M_{\odot}$
%\approx 0.3\,M_{1/2}$ 
and the power-law index $\alpha$ is set to $-1.1$ in \cite{schive2016contrasting}. 
In the above equation, $({\ud n}/{\ud m})|_\mathrm{CDM}$ is the CDM HMF.
It follows from the standard excursion-set formalism 
\cite{press1974formation,bond1991excursion,sheth1999large,sheth2001ellipsoidal} that
\begin{equation}\label{eq:generalHMF}
    \frac{\ud n}{\ud m}\bigg|_\mathrm{CDM}(m,z)
    = -\frac{\bar\rho_\mathrm{m}}{m}f(\nu)\frac{\ud \ln\sigma}{\ud m},
\end{equation}
where $\sigma(m,z)$ is the standard deviation
of the linear-theory density field at redshift $z$ smoothed on mass scale $m$,
%In this paper, $\sigma(M)$ is explicitly written as
%\begin{equation}
%\begin{aligned}
%    \sigma^2(M,z) = \int \frac{\ud k}{2\pi^2} k^2\,|W(kR)|^2 P_\mathrm{m}(k,z),\\
%    W(x)=\frac{3}{x^3}[\sin(x)-x\cos(x)],
%\end{aligned}
%\end{equation}
%where $P_\mathrm{m}(k,z)$ is the linear matter power spectrum, 
%$R$ is the length scale corresponding to the mass scale $M$, 
%$(4\pi\bar{\rho}_\mathrm{m}/3)\,R^3=M$,
%and $W(k)$ is the Fourier-space top-hat window function. 
%In Eq.~(\ref{eq:generalHMF}), 
$\nu \equiv\delta_c /\sigma(m,z)$ is the so-called peak height
($\delta_c\approx1.686$ is the comoving collapse threshold for CDM),
%usually taken to be the extrapolated linear overdensity of a spherical perturbation 
%at the time it collapses. 
%In fact, $\delta_c \simeq 1.686$ in an Einstein de Sitter cosmology (suitable for CDM)
and the function $f(\nu)$ is known as the halo multiplicity.
In this work, the functional form of $f(\nu)$ based on
the ellipsoidal collapse model
\cite{sheth2001ellipsoidal,jenkins2001mass} is adopted.
%\begin{equation}
   % f(\nu)=A\sqrt{\frac{2a}{\pi}}\,[(1+(a\nu^2)^{-p})\,\nu\,\text{exp}(-\frac{a\nu^2}{2})],
    %\label{eq:ellipsoidal}
%\end{equation}
%where the parameters $a$, $A$ and $p$ are fixed at 0.73, 0.353 and 0.175 in \verb|21cmFAST|,
%respectively, following Ref.~\cite{jenkins2001mass}.

\begin{figure}[htp]
\includegraphics[width=9cm]{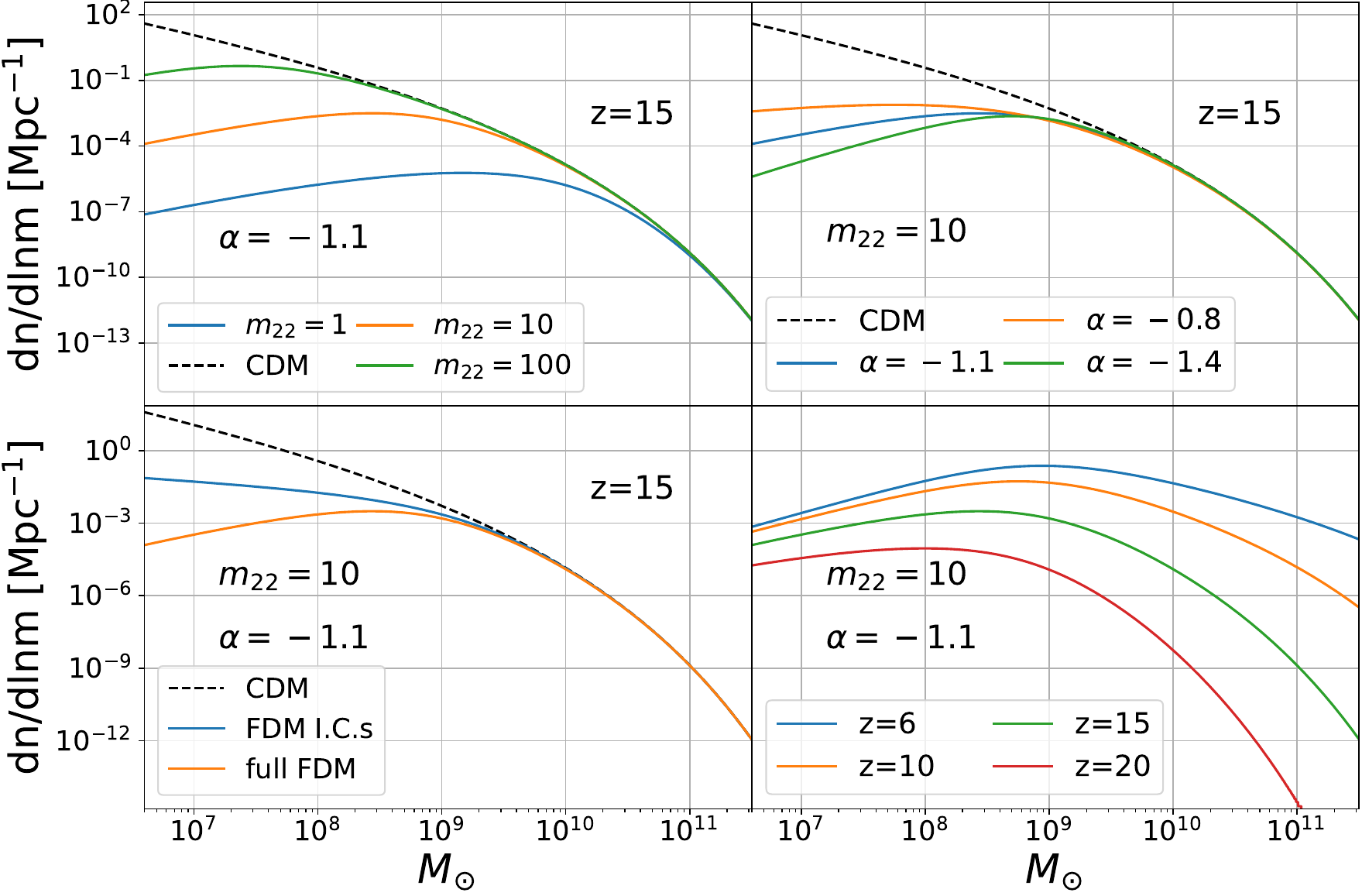}% Here is how to import EPS art
\caption{The (global) halo mass functions in different models 
for various values of parameters.\\
\emph{Upper left}: the FDM HMF for various boson masses.\\
\emph{Upper right}: the FDM HMF for various $\alpha$ indices in Eq.~(\ref{eq:HMF}).\\
\emph{Bottom left}: HMFs in different models at $z=15$.
Both the full FDM and the ``FDM I.C.s'' models assume $m_{22}=10$.\\
\emph{Bottom right}: the redshift evolution of the FDM HMF
for $m_{22}=10$ and $\alpha=-1.1$.}
\label{fig:HMF}
\end{figure}

Figure~\ref{fig:HMF} illustrates how the FDM HMF in Eq.~(\ref{eq:HMF}) changes
as the parameters of the FDM model vary.
It also shows the HMFs in different models
and the redshift evolution of the FDM HMF.
In the bottom left panel of Fig.~\ref{fig:HMF}, 
we compare the FDM HMF resulting from the full dynamics (labeled ``full FDM'') 
with the CDM case as well as an intermediate ``FDM I.C.s'' case, 
which uses the FDM linear density field as its initial conditions
but does not take into account the nonlinear dynamics of FDM.
In other words, the ``FDM I.C.s'' case assumes 
the functional form of Eq.~(\ref{eq:generalHMF}) for its HMF,
only replacing the r.m.s. of the smoothed CDM density field, $\sigma$, with that of the FDM density field
(captured by the FDM power spectrum in Eq.~[\ref{eq:ps}]).
Compared with the actual CDM HMF, the ``FDM I.C.s'' case with $m_{22}=10$
exhibits an intermediate degree of small-scale suppression, 
while the full FDM HMF exhibits the most suppression, 
as illustrated in Fig.~\ref{fig:HMF}.

Furthermore, it should be noted that the fitting formula
from Refs.~\cite{schive2016contrasting,may2021structure,may2023halo},
Eq.~(\ref{eq:HMF}), only describes the global HMF, 
an average over the entire volume.
Therefore, to describe the number of halos formed in different density environments, 
i.e., to discriminate between overdense and underdense regions,
a prescription for the density-modulated HMF is needed.
In a CDM cosmology, this is typically treated
by modifying the peak height variable, $\nu$, in Eq.~(\ref{eq:generalHMF})
\cite{mo1996analytic,cooray2002halo}.
%To consider the denser cells may be thought of as regions in which the critical density for collapse is easier to reach  
In this paper, we adopt a similar approach.
We will work with the ansatz where
the peak height variable that affects the FDM HMF in Eq.~(\ref{eq:HMF})
via the $({\ud n}/{\ud m})|_\mathrm{CDM}$ term should be written as
\begin{equation}
    \nu^2 = \frac{\left[\delta_\mathrm{c} - \delta_\mathrm{FDM}(z)\right]^2}
    {\sigma_\mathrm{CDM}^2(m,z) - \sigma_\mathrm{FDM}^2(M,z)}\,, \label{eq:ansatz}
\end{equation}
where $m$ is the halo mass, 
$M$ is the total mass within the comoving volume under consideration,
$\delta_\mathrm{FDM}(z)$ is the linear-theory FDM overdensity within this volume at redshift $z$,
and $\sigma_\mathrm{FDM}^2(M,z)$ is the variance
of the linear-theory FDM density field smoothed on mass scale $M$.
Thus, the density-modulated environmental effects are treated
using the actual FDM linear density field,
indicated by the second terms in both the numerator and the denominator above.
On very large scales ($M\to\infty$), 
the density-modulated HMF resulting from this ansatz
reduces to the global average, Eq.~(\ref{eq:HMF}), as expected.

\subsection{\label{sec:Astro}Astrophysics}

The \verb|21cmFAST| code models the reionization process
by counting the number of ionizing photons.
Here we briefly review its algorithm and the relevant astrophysics.
At each redshift, \verb|21cmFAST| tags any cell at coordinates $(\mathbf{x}, z)$ 
within the simulation volume as fully ionized 
if the number of photons per baryon exceeds unity, schematically expressed as
\begin{equation}\label{eq:ionization}
    \zeta \,f_\mathrm{coll}(\mathbf{x},z; R,{M}_\mathrm{min}) \ge 1,
\end{equation}
where $f_\mathrm{coll}(x, z; R, {M}_\mathrm{min})$
is the fraction of collapsed matter residing in halos
more massive than ${M}_\mathrm{min}$
within a comoving spherical volume of radius $R$ (the ``collapse fraction''),
and $\zeta$ is an ionizing efficiency 
that describes the conversion of mass into ionizing photons. 
In \verb|21cmFAST|, the collapse fraction is proportional to
the integral of the FDM HMF, Eq.~(\ref{eq:HMF}), 
from ${M}_\mathrm{min}$ to infinity.
We insert the environmental modulation ansatz, Eq.~(\ref{eq:ansatz}),
into the FDM HMF with $M=(4\pi \bar\rho_\mathrm{m}/3)\,R^3$,
the total mass within the comoving sphere.
%Partial ionizations are included for voxels not fully ionized by setting their ionized fractions to $\zeta f_\text{coll} (\mathbf{x},z,R,\bar{M}_{min})$.

The ionizing efficiency is implemented in two ways
in the current version of \verb|21cmFAST|: 
(i) the basic, mass-independent model
and (ii) the more realistic, mass-dependent model.
The former treats the ionizing efficiency as a single parameter, $\zeta$,
independent of the halo mass, so that the ionization algorithm
is literally expressed by Eq.~(\ref{eq:ionization}).
The latter assumes a power-law mass dependence, 
involving more free parameters to describe
the fraction of halo baryons ending up in stars, $f_*$, 
and the fraction of photons escaping into the IGM, $f_\mathrm{esc}$
(the ``escape fraction'').
Details of the mass-dependent model are described in Ref.~\cite{park2019inferring}. 
%In this model, we can still formally separate a constant``$\zeta$'' term and an %$f_{\text{coll}}$ term that includes the mass including terms,

The minimum halo mass, ${M}_\mathrm{min}$, 
is also modeled differently between the above two implementations.
In the mass-independent model, ${M}_\mathrm{min}$ is related to
the (average) minimum virial temperature of star-forming galaxies, $T_{\text{vir}}$;
see Ref.~\cite{barkana2001beginning} 
for the relationship between ${M}_\mathrm{min}$ and $T_{\text{vir}}$.
In the mass-dependent model, ${M}_\mathrm{min}$ is effectively
a free parameter of the astrophysics sector itself.
%\begin{equation}
%    \begin{split}
%        M_{\text{min}} &= 10^8 \,h^{-1} \left( \frac{\mu}{0.6} \right)^{-3/2} 
%\left( \frac{\Omega_\mathrm{m}}{\Omega_\mathrm{m}^z} \frac{\Delta_c}{18\pi^2} \right)^{-1/2}\\
%&\times \left( \frac{T_{\text{vir}}}{1.98 \times 10^4 \, \text{K}} \right)^{3/2}
%\left( \frac{1 + z}{10} \right)^{-3/2} M_{\odot} \,,
%    \end{split}
%\end{equation}
%where $\mu$ is the mean molecular weight, 
%$\Omega^z_\mathrm{m}\equiv\Omega_\mathrm{m}(1 + z)^3 /[\Omega_\mathrm{m} (1 +z)^3 + \Omega_\mathrm{\Lambda} ]$, 
%and $\Delta_c \equiv 18\pi^2 + 82d - 39d^2$ where $d = \Omega^z_\mathrm{m} - 1$.

In this work, we perform simulations
using both implementations of the \verb|21cmFAST| code.
Nonetheless, our main results are based on the mass-independent model. 
The results of the mass-dependent model
and its degeneracy with the FDM dynamics are discussed in \S\ref{sec:discussion}.

Finally, the \verb|21cmFAST| code traces
the evolution of (i) the kinetic temperature of the gas
and (ii) the angle-averaged specific intensity
of the Ly$\alpha$ radiation in each cell.
The former requires the code to compute the intensity
of the soft-band x-ray photons ($<2$ keV)
responsible for the heating of the IGM.
It is parametrized by the soft-band x-ray luminosity per star formation rate (SFR),
$L_{\text{X}< 2 \text{keV}} /\text{SFR}$
(in $\text{erg}\, \text{s}^{-1}\, \text{M}_\odot^{-1}\, \mathrm{yr}$)
\cite{park2019inferring}.
For simplicity, we refer to this parameter as $L_\mathrm{X}/\mathrm{SFR}$
in the rest of the paper.

%calculated by integrating the comoving X-ray specific emissivity $\epsilon(x,E,z')$ along the lightcone.
%\begin{equation}
   % J(\mathbf{x},E,z)=\frac{(1+z)^3}{4\pi}\int_z^{\infty} dz' \frac{cdt}{dz'} \epsilon_X e^{-\tau}
%\end{equation}
%Here, $e^{-\tau}$ corresponds to the probability that a photon emitted at an earlier time, $z'$ , survives until $z$ owing to IGM attenuation (see equation 16 of Mesinger et al. 2011) and the comoving specific emissivity is evaluated in the emitted frame, 
%$E_e = E(1 + z' )/(1 + z)$, 
%with
%\begin{equation}
    %\epsilon(\mathbf{x},E_e,z')=\frac{L_X}{SFR}[\rho_{crit,0}\Omega_b f_* (1+\delta_{nl})\frac{df_{coll}}{dt}]
%\end{equation}
%where the quantity in square brackets is the star formation rate (SFR) density along the light-cone, with $\rho_{crit,0}$ being the current critical density.\\
%Finally, the normalized X-ray efficiency is an integrated soft-band ($<2 keV$) luminosity per SFR (in $erg s^{-1} M_\odot^{-1} yr$):
%\begin{equation}
    %L_{X<2keV}/SFR=\int_{E_0}^{2keV}dEL_X/SFR
%\end{equation}

\subsection{\label{ssec:simulation}21cmFAST simulations}

In summary, we consider two FDM parameters, $(m_{22},\,\alpha)$,
and three astrophysical parameters, 
$(\zeta,\,T_\mathrm{vir},\,L_\mathrm{X}/\mathrm{SFR})$, 
in all our simulations (based on the mass-independent model; cf.~\S\ref{sec:Astro}). 
The ranges of their values are shown in Table~\ref{tab:table1}, 
along with the values in the fiducial model.
These simulations take into account both the linear and the nonlinear dynamics of FDM.
Besides the full FDM simulations, we also perform a CDM simulation
and an intermediate ``FDM I.C.s'' simulation with $m_{22}=10$ 
(described in \S\ref{sssec:FDMHMF}, similar to the \verb|21cmFAST| runs
in \cite{jones2021fuzzy}), for the sake of comparison.
Both the CDM and the ``FDM I.C.s'' simulations
adopt the same values of the astrophysical parameters
as those in the full FDM simulations.

\begin{table}[ht]%The best place to locate the table environment is directly after its first reference in text
\caption{Free parameters in our {\scriptsize 21cmFAST} simulations 
(based on the mass-independent model), as well as their fiducial values.
The CDM and the ``FDM I.C.s'' simulations also adopt these values.
In addition, we consider a range of values of $(m_{22},\alpha)$
for FDM simulations, as specified in the table.}
\begin{ruledtabular}
\begin{tabular}{cccc}
\textrm{Parameter}&
\textrm{Symbol}&
%\multicolumn{1}{c}{\textrm{value}}&
\textrm{Fiducial}&
\textrm{Range}\\
\colrule
Normalized FDM Particle Mass &$m_{22}$& 10 &(1, 100)  \\
Power-law Index in FDM HMF&$\alpha$&-1.1&(-1.4, -0.8) \\
Ionizing Efficiency &$\zeta$& 20&- \\
Minimum Virial Temperature &$T_\text{vir}$&$2\times10^4$&- \\
Soft-band x-ray luminosity &$L_\text{X}/\text{SFR}$&$10^{40}$&- \\
\end{tabular}
\end{ruledtabular}
\label{tab:table1}
\end{table}
%& $erg s^{-1} M_\odot^{-1} yr$ \\

All simulations in this work are carried out on a $L = 300\,\mathrm{Mpc}$ box,
initialized with the \emph{Planck-2018} cosmological parameters
\cite{plank2018cosmologicalparameters}, 
postprocessed according to the Zel'dovich approximation \cite{zel1970gravitational},
and down-sampled to a final resolution of $200^3$.

\section{\label{sec:results}results}

We present our main results in three parts.
In \S\ref{sec:HI}, we show the evolution of
the volume-averaged neutral fraction, $\bar x_\mathrm{H\,I}$,
for various parameters of the FDM model, $(m_{22},\alpha)$.
We discuss the global 21-cm signal in \S\ref{sec:Tb},
where we revisit the signature epochs during CD/EoR and their physical origins.
We describe the impact of FDM structure formation
on the timing and duration of these epochs.
In \S\ref{sec:ps}, we focus on the 21-cm power spectrum and its redshift evolution, 
which serves as the basis for our Fisher matrix forecasts later.

\subsection{\label{sec:HI}Average Neutral Hydrogen Fraction}

\begin{figure}[t!]
\includegraphics[width=8.5cm]{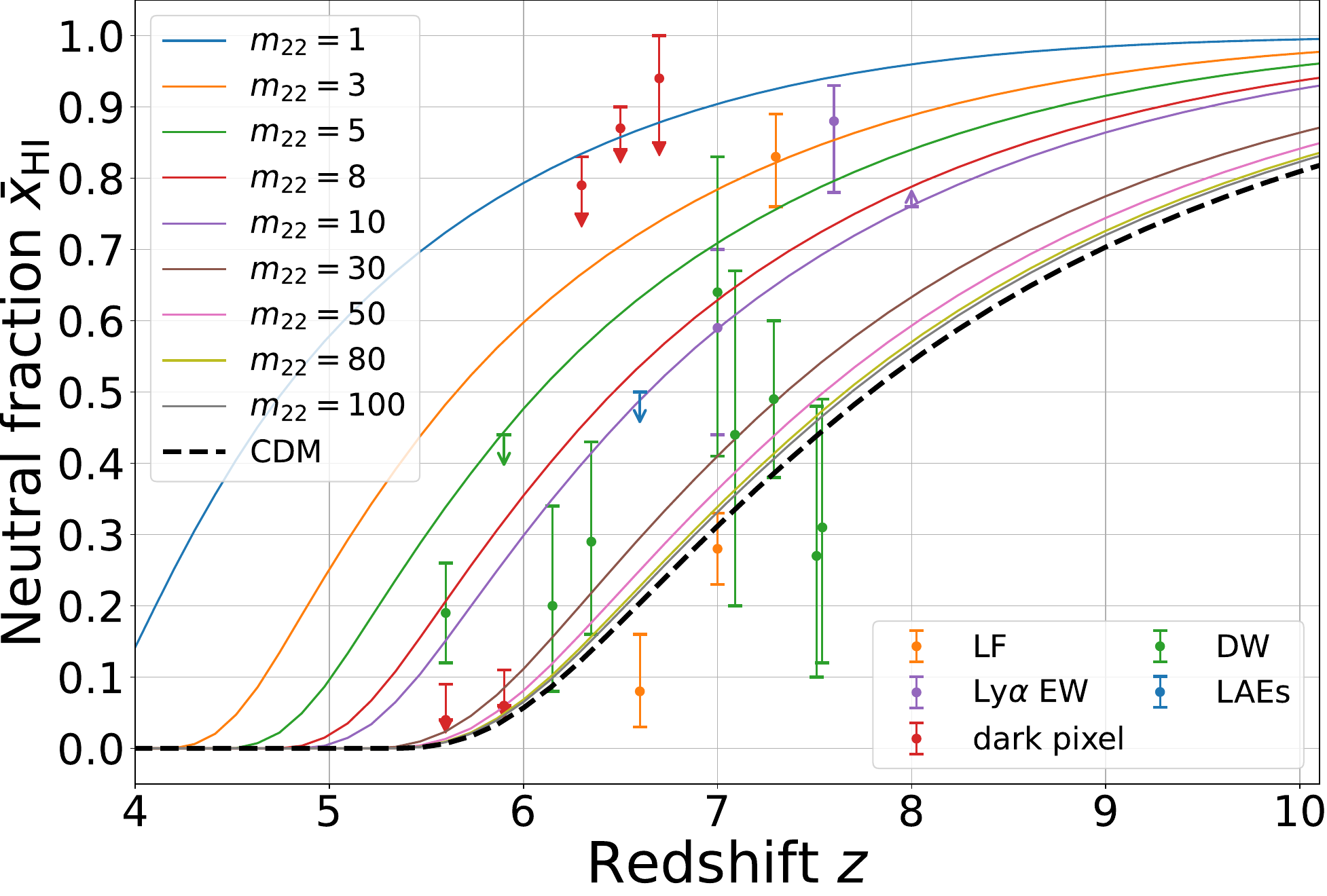}% Here is how to import EPS art
\caption{Evolution of the neutral hydrogen fraction near the end of reionization
in multiple FDM models with various boson masses and a CDM model.
All models assume $\zeta=20$, $T_\mathrm{vir}=2\times10^4$\,K,
$L_\mathrm{X}/\mathrm{SFR} = 10^{40}\,\mathrm{erg}\,\mathrm{s}^{-1}\,\mathrm{M}_\odot^{-1}\,\mathrm{yr}$
for the astrophysical parameters.
All the FDM models assume $\alpha=-1.1$ for the index parameter in the FDM HMF,
Eq.~(\ref{eq:HMF}).
The figure also displays a series of current observational constraints,
including those from measurements of the dark pixel fractions
in the Ly$\alpha$ and Ly$\beta$ forests \cite{mcgreer2015model,jin2023nearly},
clustering of Ly$\alpha$ emitters (LAEs) \cite{2010ApJ...723..869O},
the Ly$\alpha$ luminosity function (LF) \cite{morales2021evolution},
the Ly$\alpha$ equivalent widths (EWs)
\cite{mason2018universe,hoag2019constraining,mason2019inferences},
and the damping wings (DWs) in the spectra of high-$z$ quasars
\cite{greig2022igm,greig2024igm,spina2024damping}.
All the bounds shown here correspond
to the $1\sigma$ confidence intervals.}
\label{fig:xH}
\end{figure}

\begin{figure}[ht]
\includegraphics[width=8.5cm]{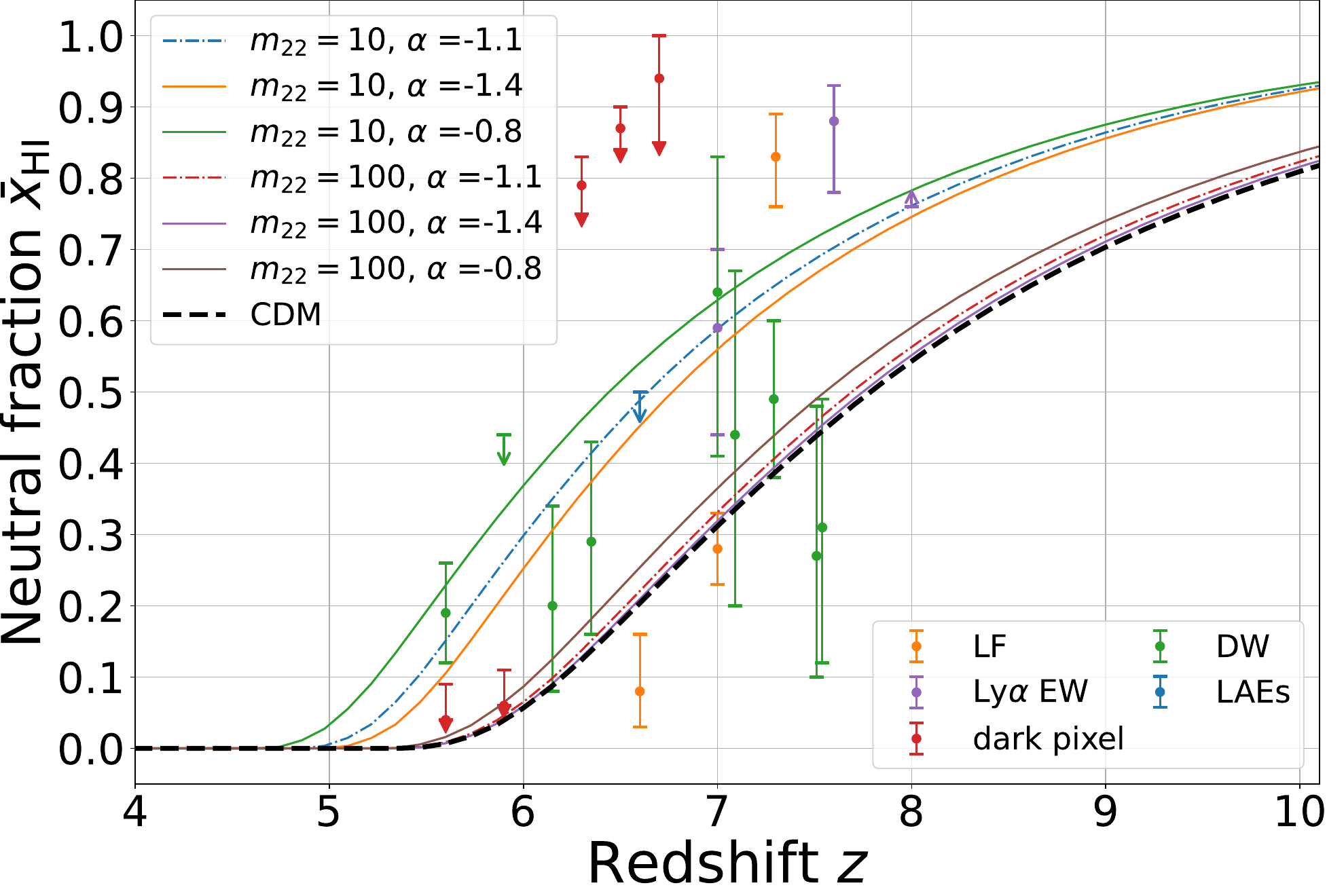}% Here is how to import EPS art
\caption{Similar to Fig.~\ref{fig:xH},
but for various $\alpha$ indices in the FDM HMF, Eq.~(\ref{eq:HMF}),
for two representative boson masses.}
\label{fig:xHin}
\end{figure}

First, we examine the ionization history in FDM cosmologies,
indicated by the volume-averaged fraction of neutral hydrogen,
$\bar x_\mathrm{H\,I}$, as a function of redshift.
We show the evolution of the neutral fraction near the end of reionization
for various boson masses in Fig.~\ref{fig:xH},
along with the existing observational constraints on $\bar x_\mathrm{H\,I}$.
As expected, reionization finishes later for smaller $m_{22}$,
since the production of ionizing photons is delayed due to the lack of dwarf halos.
On the other hand, for a boson mass as large as $m_{22}=100$,
the ionization history is so close to the CDM limit
that the FDM prediction cannot be distinguished from the CDM prediction
by current observational data.
The influence of varying the power-law index in the FDM HMF, 
i.e., the $\alpha$ parameter in Eq.~(\ref{eq:HMF}),
on the ionization history is shown in Fig.~\ref{fig:xHin}.
For fixed $m_\mathrm{{22}}$, we find that
changing the $\alpha$ index has little effect on the evolution of $x_\mathrm{H\,I}$.

\subsection{\label{sec:Tb}Global 21-cm signal}

\begin{figure*}[htbp]
\resizebox{\textwidth}{!}{\includegraphics{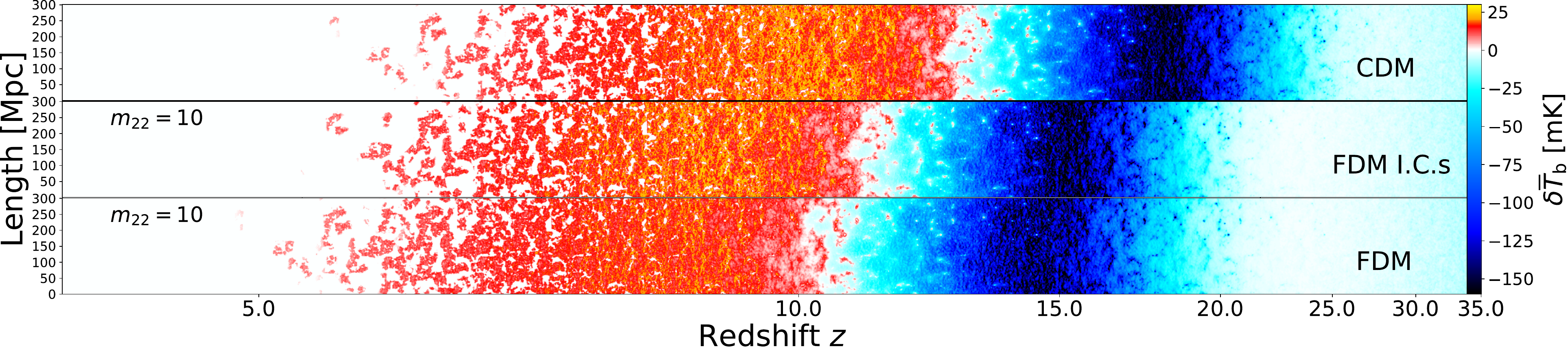}}
\caption{Evolution of the 21-cm brightness temperature over the past light cone.
The figure illustrates the comparison among the CDM model,
the intermediate ``FDM I.C.s'' model and the fiducial FDM model.
The ``FDM I.C.s'' considers the FDM linear density field as initial conditions
but does not take the full FDM nonlinear dynamics into account,
as described in \S\ref{sec:model}.
Both this hybrid case and the full FDM case assume a boson mass of $m_\text{22}=10$
and exhibit delayed CD/EoR compared with the CDM case.
The delay in the completion of reionization is strongest in the full FDM model.}
\label{fig:lcn}
\end{figure*}

The global 21-cm signal is the average 21-cm brightness temperature
throughout the entire sky as a function of redshift, $\delta\bar{T}_\mathrm{b}(z)$.
The effects of FDM structure formation on the global 21-cm signal
are illustrated in Fig.~\ref{fig:lcn}, where we display
the evolution of $\delta\bar{T}_\mathrm{b}$ over the past light cone during CD/EoR 
for three types of simulations described in \S\ref{ssec:simulation}:
the CDM simulation, the intermediate ``FDM I.C.s'' simulation, 
and the fiducial full FDM simulation with $m_{22}=10$. 
In all three cases, we can identify the two signature epochs during cosmic dawn:
the Ly$\alpha$ coupling epoch and the x-ray heating epoch,
indicated by the two transitions in Fig.~\ref{fig:lcn}.
The Ly$\alpha$ coupling epoch ($z\sim 27-18$ in the CDM universe)
refers to the process in which
the spin temperature of the H\,{\footnotesize I} gas, $T_\mathrm{S}$,
is coupled to its kinetic temperature, $T_\mathrm{K}$,
by the Ly$\alpha$ photons from the first stars, 
also known as the Wouthuysen-Field effect
\cite{wouthuysen1952excitation,field1958excitation}.
%which causes 21-cm absorption. 
The x-ray heating epoch ($z\sim 18-12$ in the CDM universe) 
results from the heating of the IGM due to the x-ray photons emitted by
early active galactic nuclei.
%reducing 21-cm absorption and eventually shifting to emission. 
%After the gas is fully heated, hydrogen reionization begins, driven by UV emissions from later generations of stars, leading the global signal to gradually decline as neutral hydrogen diminishes \cite{furlanetto2004growth}.
Figure~\ref{fig:lcn} shows that both the linear and the nonlinear FDM dynamics
contribute to the delay of these two epochs as well as the EoR. 
%and the delay is strongest in the full FDM universe.

\begin{figure}[ht]
\includegraphics[width=9cm]{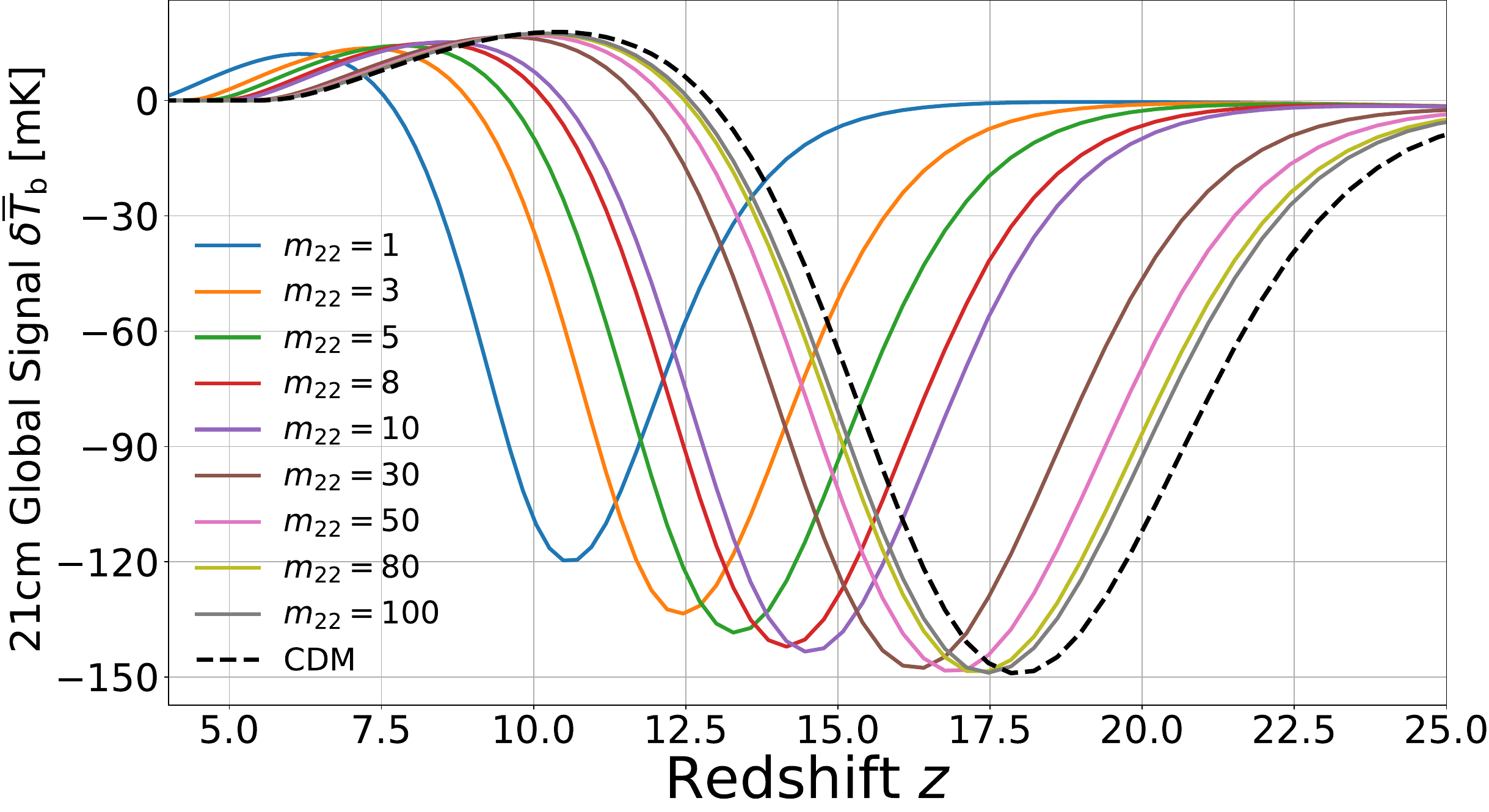}% Here is how to import EPS art
\caption{Evolution of the global 21-cm signal 
in multiple FDM models with various boson masses and a CDM model.
These models have the same parameter values as in Fig.~\ref{fig:xH}.}
\label{fig:temp}
\end{figure}

\begin{figure}[ht]
\includegraphics[width=9cm]{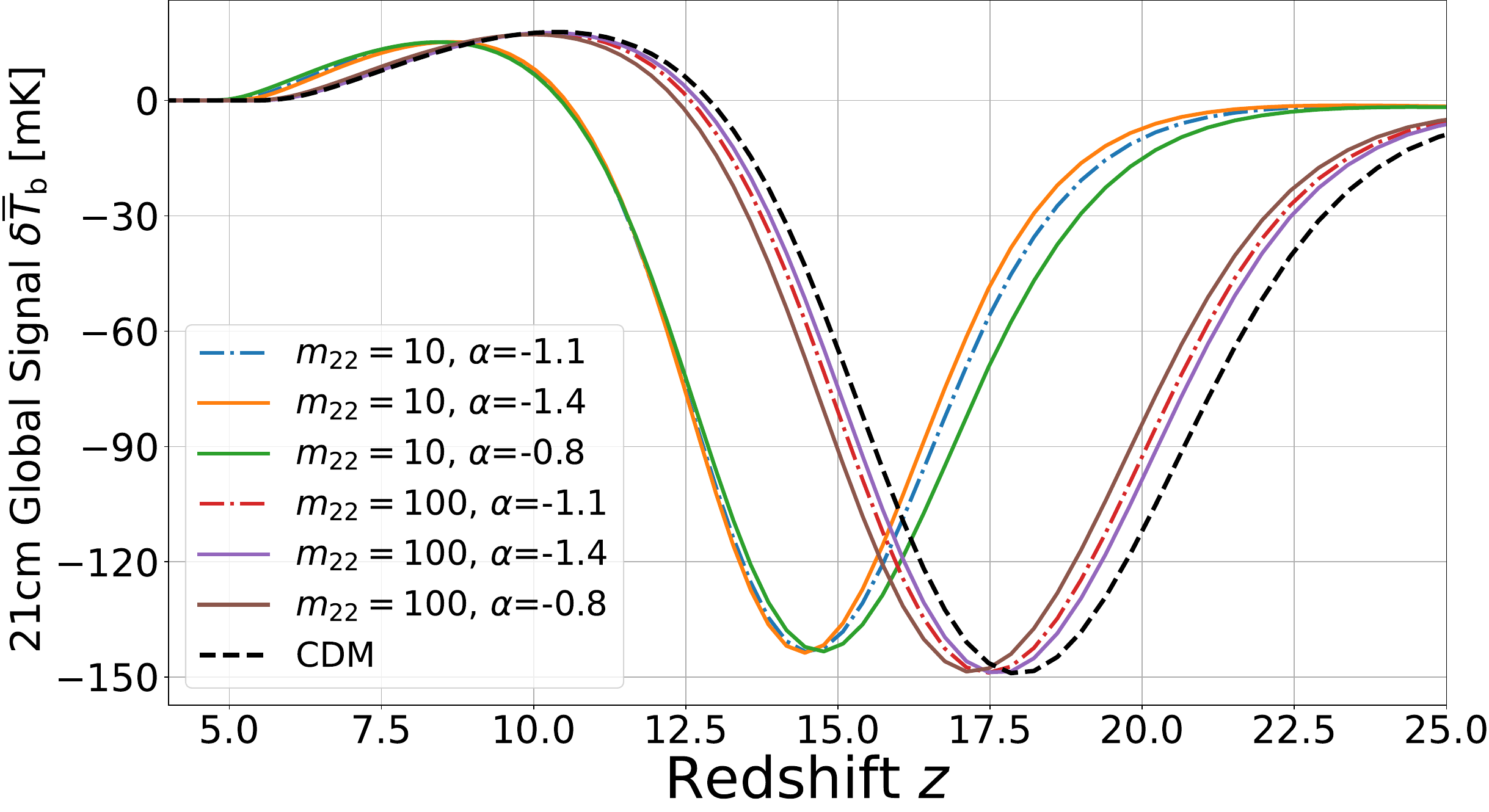}% Here is how to import EPS art
\caption{Similar to Fig.~\ref{fig:temp},
but for various $\alpha$ indices in the FDM HMF, Eq.~(\ref{eq:HMF}),
for two representative boson masses.
These models have the same parameter values as in Fig.~\ref{fig:xHin}.}
\label{fig:tempin}
\end{figure}

The quantitative evolution of the global 21-cm signal, $\delta\bar{T}_\mathrm{b}(z)$,
in full FDM simulations is illustrated in Fig.~\ref{fig:temp}.
For comparison, we show a series of models with various boson masses
along with our fiducial model (cf.~Table~\ref{tab:table1});
these models are the same as in Fig.~\ref{fig:xH}.
We find again that the delaying effect is stronger for smaller $m_\text{{22}}$, 
indicated by the locations of the minima, the maxima,
and the zero-crossing points of the $\delta\bar{T}_\mathrm{b}(z)$ curves. 
Meanwhile, Fig.~\ref{fig:temp} also shows that
the duration of each epoch is shortened compared with the CDM model.
%slightly mitigating the delaying effect.
This is indicated, for example,
by the increasing minimum values of $\delta\bar{T}_\mathrm{b}(z)$
that correspond to the onset of x-ray heating, as $m_{22}$ decreases.
The reason for this shortening is that both the number of x-ray photons
and that of ionizing photons depend directly on the collapse fraction, 
and the latter is proportional to the integral of the HMF
whose high-mass end is unaffected by the FDM dynamics, 
thereby mitigating the delaying effect.

The influence of varying the $\alpha$ index in the FDM HMF
on the global 21-cm signal is shown in Fig.~\ref{fig:tempin}. 
Again, changing the $\alpha$ index
has little effect on the global signal for fixed $m_{22}$.

\begin{figure*}[htb]
\includegraphics[width=18cm]{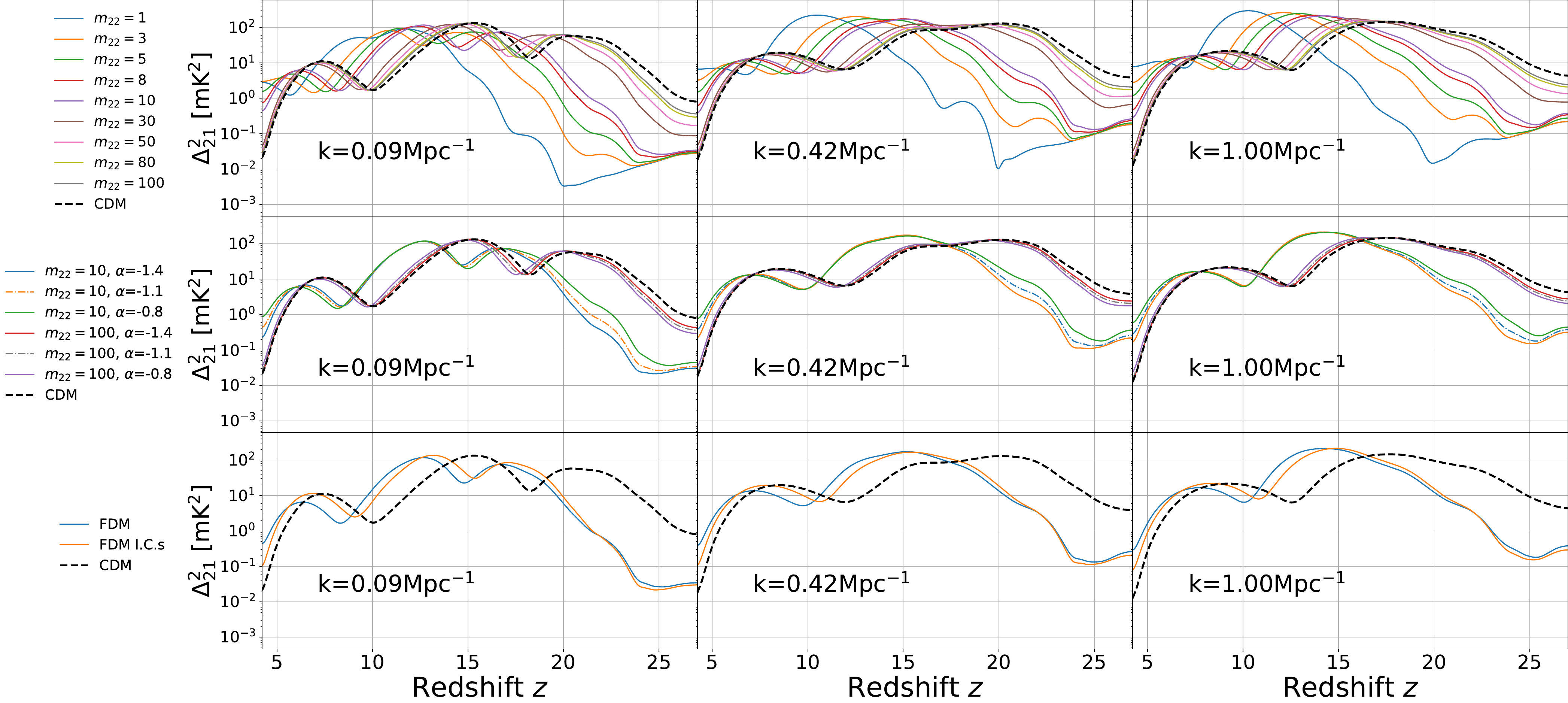}% Here is how to import EPS art
\caption{Evolutions of the dimensionless 21-cm power
at three different wave numbers during CD/EoR
in various FDM universes and a CDM universe. 
The \emph{top row} displays the effect of varying the boson mass. 
%The height of the Ly$\alpha$ coupling peak is increase and decrease with the increase wavenumber $k$, and the X-ray heating peak is decrease and increase with the increase wavenumber $k$. \\
The \emph{middle row} displays the effect of varying the $\alpha$ index in the FDM HMF;
see Eq.~(\ref{eq:HMF}).
The \emph{bottom row} compares the $\Delta^2_{21}(k)$ evolution
in three different models; see \S\ref{ssec:simulation}.
Both the full FDM and the ``FDM I.C.s'' models assume $m_{22}=10$.}
\label{fig:pstot}
\end{figure*}

%First, the Lyman-$\alpha$ coupling era, lasting from $z \sim 27–18$, from the first stellar formation until the minimum of $T_{21}$, when X-ray heating starts to dominate. During this era the first stars emit UV photons which, after redshifting into the Lyman-$\alpha$ transition, produce Wouthuysen-Field coupling between the hydrogen spin and kinetic temperature, giving rise to (anisotropic) 21-cm absorption.

%Second, the epoch of heating, which starts at $z \simeq 18$ and does not end until $z \simeq 12.5$, at which point the 21cm signal crosses zero. In this epoch the x-rays produced in the first galaxies heat up the hydrogen, inhomogeneously reducing the amount of 21-cm absorption, until it turns into emission. After the gas is fully heated, the hydrogen will start getting reionized by the UV emission from subsequent generations of stars. This marks the beginning of the EoR, during which the global signal smoothly decays to zero as the neutral hydrogen fraction slowly vanishes \cite{furlanetto2004growth,mcquinn2006cosmological}.

\subsection{\label{sec:ps}21-cm power spectrum}

21-cm experiments based on radio interferometers
mainly observe the fluctuations of the 21-cm brightness temperature in Fourier space.
These fluctuations are statistically described
by the 21-cm power spectrum, $P_{21}(k)$, defined as
\begin{equation}
    \langle \delta T^*_\mathrm{b}(\vec k)\,\delta T_\mathrm{b}(\vec{k}') \rangle 
    \equiv (2\pi)^3\,P_{21}(k)\,\delta^{(3)}_\mathrm{D}(\vec k - \vec k'),
\end{equation}
where $\delta^{(3)}_\mathrm{D}$ denotes the Dirac delta function.
Throughout the paper, we work with the dimensionless power spectrum,
$\Delta^2_{21}(k) \equiv k^3 P_{21}(k)/(2\pi^2)$, in units of $(\mathrm{mK})^2$.
$\Delta^2_{21}(k)$ describes the variance of
the 21-cm brightness temperature fluctuations per logarithmic wave number.

%\begin{figure}
%\includegraphics[width=8.5cm]{ps-z_ind_log.pdf}% Here is how to import EPS art
%\caption{\label{fig:psind}
%Comparison between the 21-cm power spectrum evolution with redshift at fixed two wavenumbers in our fiducial CDM and two different FDM particle mass with three different HMF index $\alpha$ universe. The influence is weak, but the form of influence on the evolution of power spectrum is consistent with the previous discussion.}
%\end{figure}

\begin{figure}[ht]
\includegraphics[width=8.5cm]{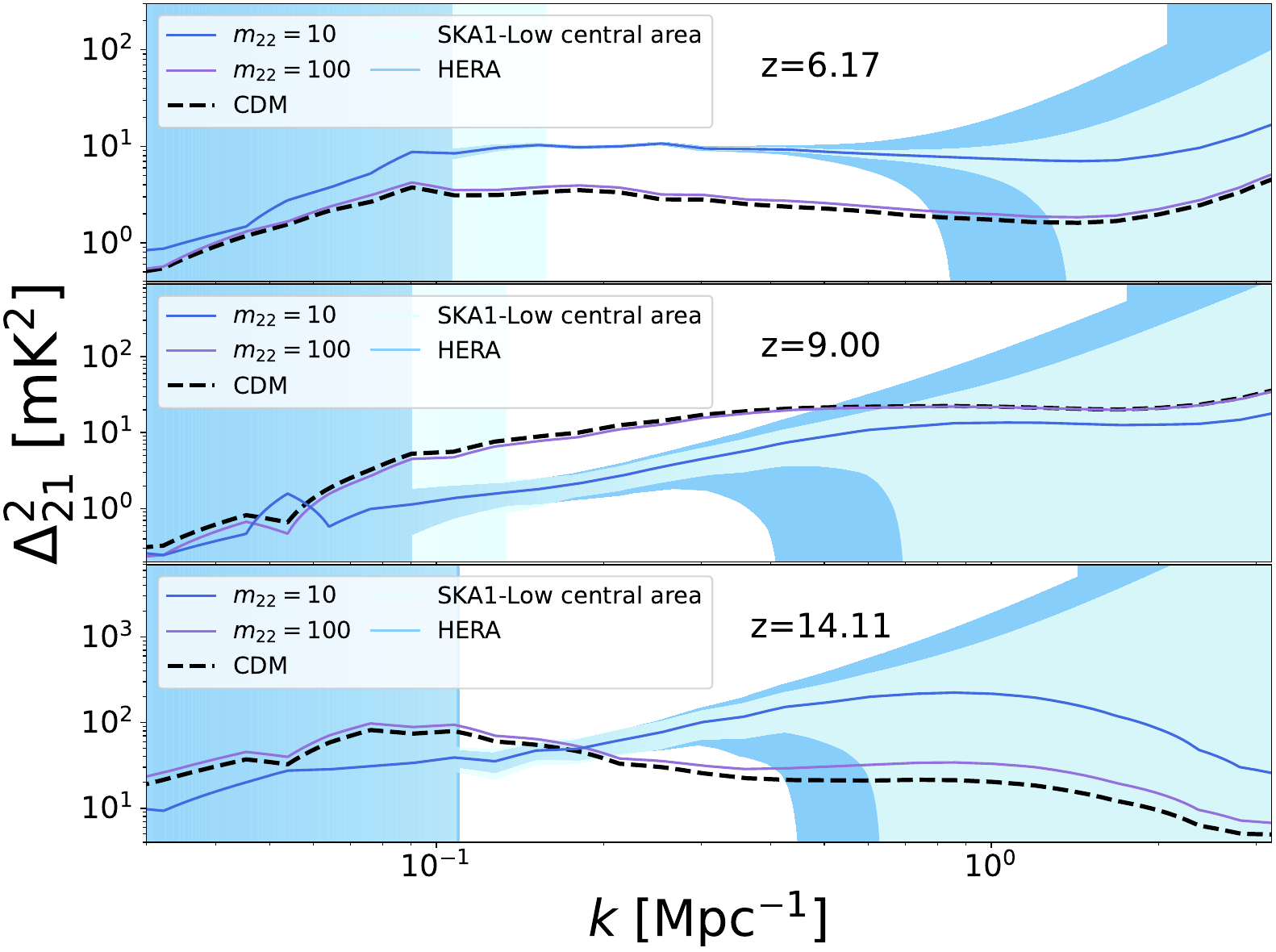}% Here is how to import EPS art
\caption{The dimensionless 21-cm power spectra at three different redshifts
in two FDM universes and a CDM universe.
The light and dark blue shaded regions represent
the expected 1$\sigma$ uncertainties of the power-spectrum measurements
by SKA1-Low (central area) and HERA, respectively,
for the fiducial FDM model ($m_{22}=10$).
For visualization purposes, the light blue shade is placed
on top of the dark blue shade at the high-$k$ end
but the layering is reversed on the low-$k$ end, 
with a smooth transition in between.}
\label{fig:psk}
\end{figure}

We illustrate the evolution of the dimensionless 21-cm power
at three different wave numbers in Fig.~\ref{fig:pstot}, 
for both the FDM and CDM cosmologies. 
The delaying effect resulting from the FDM dynamics is again significant here.
At a given wave number, all the signatures in the evolution of the 21-cm power
are shifted toward lower redshifts,
%On large scales ($k \sim 0.1$ $\mathrm{Mpc}^{-1}$), 
%the evolution of the 21-cm power spectrum in the CDM universe
including the three bumps attributed to the fluctuations
in the Ly$\alpha$ coupling, x-ray heating, and ionized fractions, respectively.
Between these bumps, the negative contribution
of the cross power between those fluctuation fields 
gives rise to the two troughs in the $\Delta^2_{21}(k)$ evolution
\cite{2008ApJ...680..962L,mesinger2013signatures}, dubbed ``early'' and ``late''.

%and the Ly-$\alpha$ coupling bump height lower than X-ray heating peak. 
%At small scales ($k>0.4\,\mathrm{Mpc}^{-1}$), however, this negative contribution become weaker, the height of the Ly$\alpha$ coupling bump decreases firstly and than rises as the scale smaller, but X-ray heating peak is the opposite, and the trough almost disappear, so the power in high redshift is larger overall.

Apart from the overall delaying effect,
the top row of Fig.~\ref{fig:pstot} also shows that
the early trough between the Ly$\alpha$ coupling epoch and the x-ray heating epoch
in the evolution of 21-cm powers diminishes for small boson masses ($m_{22}<10$).
Since this trough corresponds to the stage
in which $T_\mathrm{S}$ is well coupled to $T_\mathrm{K}$
for all H\,{\footnotesize I} gas \cite{jones2021fuzzy},
its diminishing suggests that
when the FDM suppression of small-scale structure formation is strong,
the deficiency in Ly$\alpha$ photons from the first stars is so substantial that
there is not sufficient time for Ly$\alpha$ coupling to complete for all H\,{\footnotesize I} gas
before the first x-ray sources form and heat the IGM.
The rise of $T_\mathrm{K}$ then
increases the fluctuations in $\delta T_\mathrm{b}(\vec x)$,
boosting the 21-cm power and yielding the x-ray heating bump in $\Delta^2_{21}(k,z)$.
On nonlinear scales ($k>0.1\,\mathrm{Mpc}^{-1}$),
the trough essentially disappears for small $m_{22}$
and the Ly$\alpha$ coupling bump merges with the x-ray heating bump.

For fixed $m_{22}$, changing the $\alpha$ index in the FDM HMF
does not lead to measurable differences in the $\Delta^2_{21}(k)$ evolution,
as illustrated in the middle row of Fig.~\ref{fig:pstot}.
For the comparison between the full FDM and the intermediate ``FDM I.C.s'' models,
their differences do not appear to be considerable until x-ray heating begins,
as indicated by the bottom row of Fig.~\ref{fig:pstot}.
The reason is that at higher redshifts ($z\gtrsim 15$),
the HMFs in these two cases only differ on the smallest mass scales
where (mini)halos do not host star formation \footnote{We do not
consider molecular cooling in this work.}.
We can further infer that the FDM effects on CD/EoR
are dominated by its linear dynamics at early times,
while the nonlinear wave dynamics becomes important at late times.

Figure~\ref{fig:psk} presents the dimensionless 21-cm power spectra at given redshifts. 
The shape of $\Delta_{21}^2(k)$ in the FDM case is similar to that in the CDM case,
remaining mostly featureless.
The relative amplitudes between the fiducial FDM and the CDM cases
shown in the top and middle panels of Fig.~\ref{fig:psk}
reflect the delaying effect from FDM dynamics.
In the bottom panel, the amplitude of $\Delta_{21}^2(k)$ in the fiducial FDM universe
is smaller than that in the CDM universe on large scales
but exceeds the latter on small scales.
To interpret this boost in the small-scale power,
we note that x-ray heating has just begun in the FDM universe
by this redshift ($z=14.11$),
whereas the CDM universe is already deep into the x-ray heating epoch.
%at the beginning of X-ray heating, 
Hence, only the surroundings of the x-ray sources are heated in the FDM case
so that the fluctuations in $\delta T_\mathrm{b}$ are first enhanced on small scales.
By contrast, the $\delta T_\mathrm{b}$ fluctuations are smaller in the CDM case
since all the IGM is well heated.

In summary, all these characteristics of $\Delta_{21}^2(k,z)$ from FDM cosmologies
in both time and spatial dimensions
will potentially allow for discrimination between the FDM model and the CDM model
by radio interferometric observations,
which we discuss in the rest of the paper.

Before we proceed, it is useful to remark that
our results of the global 21-cm signal and the 21-cm power spectrum here
can be compared with those from other alternative dark matter scenarios,
such as warm dark matter, interacting dark matter with late kinetic decoupling, etc.
\cite{2014MNRAS.438.2664S,2018PhRvD..98f3021S,2019PhRvD..99b3522L,
2020PhRvD.101f3526M,2024PhRvD.110j3533V,2025JCAP...01..071P}.
Since all these models suppress the linear matter power spectrum as does FDM,
it might be difficult to discriminate between their quantitative differences in $P(k,z)$ through a model selection analysis.
However, the unique nonlinear wave dynamics of FDM (distinct from the N-body dynamics)
may leave distinguishable signatures on the halo statistics and hence the 21-cm observables at late times during the EoR.
We will examine this possibility in follow-up studies.

\section{\label{sec:forecast}observational FORECAST}

In this section, we forecast the prospect of measuring the parameters of the FDM model
using 21-cm power spectrum data from the upcoming SKA Observatory,
particularly the SKA1-Low telescope \cite{2013ExA....36..235M,bacon2020cosmology}.
We estimate the sensitivity of the power spectrum measurement
and predict the constraints on the FDM and astrophysical model parameters
using the Fisher matrix formalism.
For comparison, we also show the forecast results for HERA.

\subsection{Observation}

We consider two telescope configurations
based on SKA1-Low \cite{bacon2020cosmology,2022PASA...39...15S}
and HERA \cite{pober2013baryon,pober2014next}, 
to quantify the prospects of measuring the 21-cm power spectrum
and the ability of these experiments to detect the signatures of FDM.
The SKA1-Low interferometer array measures the 21-cm temperature fluctuations
from cosmic dawn (50 MHz or $z \sim 27$)
to the postreionization era (350 MHz or $z \sim 3$). 
The \emph{central area} of SKA1-Low contains 296 stations of 40\,m diameter,
each consisting of 256 dipole antennae.
%This central area provides most of the sensitivity to the low-$k$ modes.
We will only consider this central area for our SKA1-Low forecasts throughout the paper.
For the HERA configuration, we adopt its original core design
with 331 dishes of 14\,m diameter packed in a hexagon \cite{dillon2016redundant}.
We do not take into account the ``outrigger'' antennae, either.
%employs 350 dishes of 14\,m diameter
%(320 hexagonally packed core + 30 extended baseline antennas)
HERA covers the $50$--$250$ MHz frequency range ($z \sim 5$--$27$).
%and its full Phase II system covers the $50$--$250$ MHz frequency range ($z \sim 5$--$27$) \cite{2023ApJ...945..124H}. 
Both interferometers perform drift-scan observations
to map large sky areas using Earth rotation synthesis.

We use the open-source Python package \verb|21cmSense| \cite{pober2014next}
to model the above two telescope configurations and
estimate the sensitivity of each experiment to the 21-cm power spectrum.
%\sh{Especially, we generate the “concept HERA” in this work using the code. The concept version of HERA is a perfect hexagon (with no split-core), with 331 elements and no outriggers.}
This code accounts for the $uv$ sensitivity of each dish/station in the array
and calculates the possible errors in the 21-cm power spectrum measurement,
which includes the cosmic variance.
After determining the total \emph{coherent} observing time $t(\vec{k})$
for each $(u,v,\eta)$ cell in the Fourier space \footnote{We adopt the standard notations
for the Fourier-space coordinates $(u,v,\eta)$ from radio astronomy,
where $\vec u\equiv (u,v)$ describes the transverse modes ($k_\perp$)
and $\eta$ the longitudinal modes ($k_\parallel$).
Particularly, $\vec u=\vec b/\lambda$, where $\vec b$ is the baseline
and $\lambda$ is the observing wavelength.
The $\eta$ coordinate is the Fourier dual to the observing frequency,
defined with respect to each observational band (redshift bin).},
\verb|21cmSense| computes the variance of the optimal estimator
of the 21-cm power in this cell \cite{pober2014next,ewall2016constraining,liu2020data}.
It is expressed as
\begin{equation}\label{eq:var}
    \mathrm{var}\,[\Delta_{21}^2(\vec k, z)]
    = \left[X^2\,Y \frac{k^3}{2\pi^2}
    \frac{\Omega'}{2t(\vec k)}T_{\mathrm{sys}}^2
    + \Delta^2_{21}(k,z)\right]^2,
\end{equation}
where $X$ ($Y$) is the redshift-dependent conversion factor
between angle (frequency) and comoving distance \cite{2012ApJ...753...81P},
and $\Omega'$ is the ``effective beam area''
which describes the primary beam field of view
(cf.~Eq.~[B12] in Ref.~\cite{parsons2014new}).
%Specifically, it is the integral of the primary beam squared over solid angle divided by the solid-angle integral of the primary beam \cite{parsons2014new}, 
The system temperature, $T_\mathrm{sys}$, describes the thermal noise of the instrument.
It equals the sum of the receiver temperature, $T_\mathrm{rcv}$,
and the sky brightness temperature, $T_\mathrm{sky}$.
The dimensionless 21-cm power spectrum in Eq.~(\ref{eq:var}), $\Delta^2_{21}(k,z)$,
accounts for the sample variance.
%comes from the cosmic variance.
%In our estimate of the SKA1-Low and HERA sensitivities, 
Moreover, Eq.~(\ref{eq:var}) only applies to
coherent observations within the field of view of the drift-scan telescope.
The variance can thus be further reduced by a factor
of the number of fields on the sky
corresponding to \emph{incoherent/independent} observations
during a day as Earth rotates.
Finally, we assume a 6\,hr observing time per day and
a total observation period of 180 days, for both telescope configurations,
SKA1-Low (central area) and HERA.
We also assume a uniform bandwidth of 8\,MHz
%The 21-cm power spectra are simultaneously measured with a bandwidth of 8\,MHz
across the entire observing frequency range.
%($50-\sh{282?}350$ MHz for SKA1-Low).
%It may be unrealistic to assume that simultaneous measurements are feasible across this large frequency band, but this should nevertheless provide a useful forecast.
Other telescope parameters are summarized in Table~\ref{tab:table2}.

\begin{table*}[ht]
\caption{Design parameters of the interferometers
and user-defined observation parameters, 
required by the {\footnotesize 21cmSense} code.
The sources of the SKA1-Low parameters are specified.
The HERA parameters are provided by {\footnotesize 21cmSense}.
The bandwidth, the observing time per day,
and the total number of days in the observation are user defined.
The bandwidth parameter determines the LoS comoving lengths of the survey cubes
at different (central) redshifts/frequencies.}
%\texttt{table*} environment. It also demonstates the use of
%\textbackslash\texttt{multicolumn} in rows with entries that span
%more than one column.}
\begin{ruledtabular}
\begin{tabular}{ccc}
% &\multicolumn{2}{c}{$D_{4h}^1$}&\multicolumn{2}{c}{$D_{4h}^5$}\\
 Parameters & HERA & SKA1-Low (central area) \\ \hline
 Number of dishes/stations & 331 & 296 (Ref.~\cite{skaoSciencePerf}) \\
 Observing frequency range (MHz) & $50$--$350$ & $50$--$250$\\
 Sky temperature $T_\mathrm{sky}$ (K) & $60\,({\nu}/{300\,\mathrm{MHz}})^{-2.55}$
 & $20\,({\nu}/{408\,\mathrm{MHz}})^{-2.75}$ (Ref.~\cite{skaoConfigCoords})\\
 Receiver temperature $T_\mathrm{rcv}$ (K) & 100 & 30 (Ref.~\cite{2022PASA...39...15S})\\
 Dish size (m) & 14 & 40 (Ref.~\cite{skaoConfigCoords})\\
 Frequency resolution (kHz) & 97.65 & 5.4 \footnote{This is the standard
 frequency resolution reported in the SKA1 Design Baseline document
 (SKA-TEL-SKO-0001075). For the forecasts in this paper, however,
 we take a coarser frequency resolution of 32\,kHz 
 (corresponding to 250 channels within a bandwidth of 8\,MHz)
 so that the maximum longitudinal wave number $k_\parallel$
 probed by the SKA1-Low observation is about 3.4\,Mpc$^{-1}$, 
 matching the resolution of our simulated light cones.} \\
 Bandwidth (MHz) & 8 & 8 \\
 %Integration time $t_\mathrm{{int}}$ [s] & 60 & 0.9 \footnotemark[3] \\
 Observing time per day (hrs) & 6 & 6 \\
 Number of days & 180 & 180
\end{tabular}
\end{ruledtabular}
\label{tab:table2}
\end{table*}

In addition, the \verb|21cmSense| code takes into account
foreground contamination to the 21-cm power spectrum measurements,
which mostly affects the Fourier modes located in a ``wedge''
in the $k_\perp-k_\parallel$ space \cite{2012ApJ...752..137M}.
Specifically, the modes in the foreground wedge are masked out
in the calculation of the power-spectrum sensitivity.
However, it is still unclear where to define the edge of this wedge.
In the implementation of the \verb|21cmSense| code,
three models for the foreground wedge are considered,
termed ``pessimistic, moderate and optimistic'' \cite{pober2014next}.
We apply the moderate model to all cases in this work.

The resultant estimated errors of the 21-cm power spectrum measurements
are illustrated in Fig.~\ref{fig:psk}.
SKA1-Low (central area) has a slightly broader sensitive range of wave numbers than HERA
but both are roughly within $[0.1,1]\,\mathrm{Mpc}^{-1}$.
Large-scale sensitivities are primarily provided by $k_\parallel$,
because the survey volume of each band is basically an elongated rectilinear cube along the LoS.
On top of that, the largest scale (minimum $k$) of the sensitivity range
is further determined by the foreground wedge of each interferometer.
This explains the difference in values of $k_\mathrm{min}$
between HERA and SKA1-Low, as shown in the top and middle panels of Fig.~\ref{fig:psk}.
On small scales, the measurement errors of $\Delta_{21}^2(\vec k)$
are dominated by the thermal noise,
where SKA1-Low displays an advantage over HERA.

\subsection{Fisher Matrix Forecasts}

Here we present our forecasts for the constraints on the FDM parameters
from the 21-cm power spectrum measurements by SKA1-Low (central area) and HERA,
based on the Fisher matrix formalism \cite{tegmark1997karhunen}.
The free parameters and their fiducial values are specified in Table~\ref{tab:table1}.
In the following, we consider a vector of normalized model parameters, $\vec p$,
%=(p_\zeta, {T_\mathrm{vir}}, {L_\mathrm{X}}, \alpha, {m_{22}})$.
in which each parameter is normalized by its fiducial value,
e.g., $p_{m_{22}}=m_{22}/m_{22,\mathrm{fid}}$.
%Our ﬁducial parameter value sets are ($\zeta = 20$, $T_{vir} = 2\times10^4$, $L_\text{X}/\text{SFR}=10^{40}$, $\alpha=-1.1$, $m_{22}=10$). 
%Note that the fiducial models here are FDM cases with two FDM particles, rather than CDM cases, 
%since this facilitates the Fisher matrix computations,
The Fisher matrix can then be written as
\cite[e.g.,][]{pritchard2010constraining,liu2016constraining}
\begin{equation}\label{eq:fisher}
    F_{ij} = \sum_{k, z} \frac{\partial \Delta_{21}^2(k, z)}{\partial p_i} \frac{\partial \Delta_{21}^2(k, z)}{\partial p_j} 
    \frac{1}{\mathrm{var}\,[\Delta_{21}^2(k, z)]} \,,
\end{equation}
where the sum runs over all wave number bins
and the full redshift range of the observation,
and the variance of the 21-cm power spectrum is estimated
using the \verb|21cmSense| code as described above.
Each redshift bin corresponds to a survey volume of fixed bandwidth (8\,MHz).
The $z$ bins start from the minimum observing frequency, 50\,MHz for both telescopes.
For each $z$ bin, the wave number bins are given by the output of \verb|21cmSense|.
They are essentially determined by $k_\parallel$
but also affected by the foreground wedge, as mentioned above.
%In detail, we divide the light cone evolving along the redshift into several chunks according to the bandwidth 8\,MHz start at 50\,MHz. For SKA1-Low, the end frequency is 282\,MHz and for the HERA is 250\,MHz. 

\begin{figure}[ht]
\includegraphics[width=7cm]{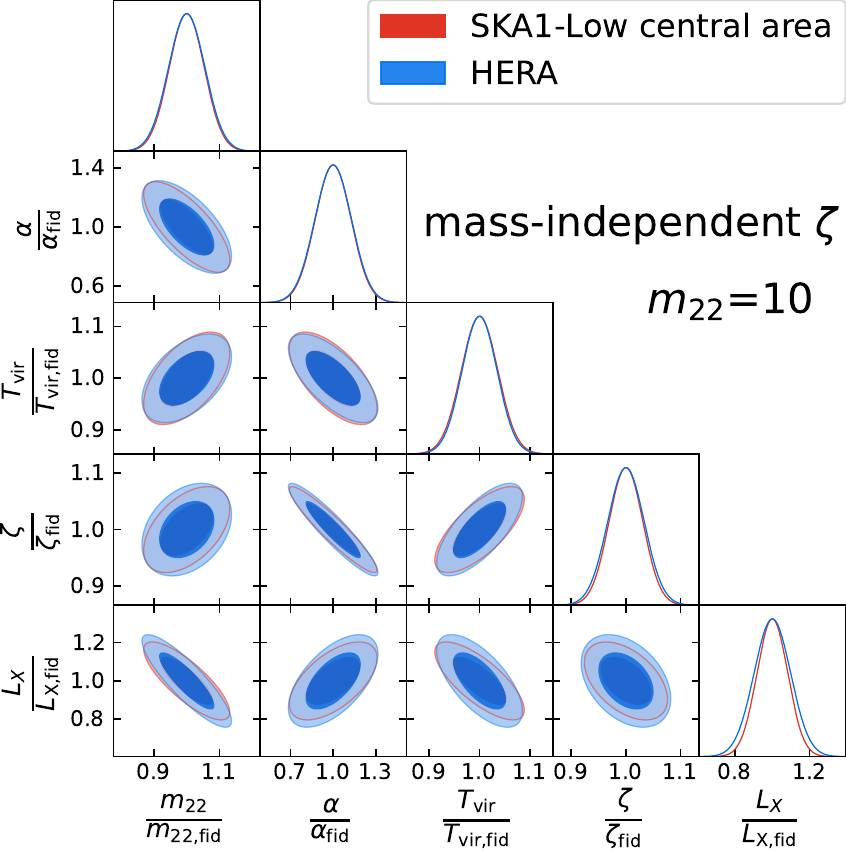}
%\subfigure{
%\includegraphics[width=7cm]{nozeta50tot.pdf}
\caption{Predicted constraints on the normalized model parameters
from the 21-cm power spectra measured by SKA1-Low (central area) and HERA, respectively,
based on the fiducial mass-independent model specified in Table~\ref{tab:table1}.
Contours contain 68\% and 95\% of the probability.}
\label{fig:nz}
\end{figure}

The inverse of the Fisher matrix, $(F^{-1})_{ij}$,
yields the covariance matrix of the (normalized) parameters.
The results are presented in Fig.~\ref{fig:nz},
which illustrates the sensitivities of SKA1-Low (central area) and HERA
to the fiducial FDM model ($m_{22}=10$). 

%\begin{table}%The best place to locate the table environment is directly after its first reference in text
%\caption{\label{tab:table3}
%The parameters in our fiducial models for Fisher matrix forecast, note %here are two fiducial models with two different FDM particle mass to %forecast the parameter constraint ability in weak and strong FDM effect universe.}
%\begin{ruledtabular}
%\begin{tabular}{ccc}
%\textrm{Parameter}&
%\textrm{Symbol}&
%\multicolumn{1}{c}{\textrm{value}}&
%\textrm{value}\\
%\colrule
%Normalized FDM Particle Mass &$m_{22}$&10, 50  \\
%FDM HMF Index&$\alpha$&-1.1 \\
%Ionizing Efficiency &$\zeta$& 20  \\
%Minimum Virial Temperature &$T_\text{vir}$&$2\times10^4$ \\
%Normalized X-ray Efficiency &$L_\text{X}$&$10^{40}$ \\ 
%\end{tabular}
%\end{ruledtabular}
%\end{table}

%We can see that the constraint ability of SKA1 is weaker than that of HERA-like, because although the dish area of SKA1-Low station is larger than that of HERA-like. However, the SKA1 center has only 296 stations spread over 3km in diameter, most of which are located at 1km in diameter (SKA1-Low core), while HERA-like is more concentrated and dense, thus performing better a small $k$ (large scale) than the SKA1-Low, and this work focuses on the analysis of large-scale structures. 

Measurement uncertainties of the (normalized) parameters
by both SKA1-Low (central area) and HERA
are shown in Table~\ref{tab:table4}.
%These results sum over the full redshift range spanned by HERA ($z$ = 5-27) , SKA1-Low central area ($z$ = 3-27) and over all wavenumbers.
%We forecast very tight constraints on the FDM mass for moderate foreground treatments in low FDM particle mass universe but not in high one:
We find that SKA1-Low/HERA should be able to constrain the boson mass
within 10.7\%/11.2\% of the fiducial value ($m_{22}=10$)
at $2\sigma$ (95\%) confidence. 

\begin{table}[ht]%The best place to locate the table environment is directly after its first reference in text
\caption{Predicted measurement uncertainties of the model parameters
by SKA1-Low (central area) and HERA, respectively.
The table shows the 1$\sigma$ confidence limits of the relative errors
with respect to the fiducial mass-independent model.
}
\begin{ruledtabular}
\begin{tabular}{cccccc}
\textrm{Parameter}&
\textrm{value}&
%\multicolumn{1}{c}{\textrm{value}}&
\textrm{SKA1-Low central area (1$\sigma$)}&
\textrm{HERA (1$\sigma$)}&
%\textrm{SKA for caseII(1$\sigma$)}&
%\textrm{HERA for caseII(1$\sigma$)}
\\
\colrule
$m_{22}$&10 & $5.37\%$ &$5.59\%$
%&$47.84\%$&$36.14\%$  
\\
$\alpha$&-1.1& $12.59\%$&$12.86\%$
%&$47.04\%$&$33.90\%$
\\
$\zeta$& 20 &$3.10\%$ &$3.35\%$
%&$5.79\%$&$4.37\%$
\\
$T_\text{vir}$&$2\times10^4$& $3.64\%$&$3.48\%$
%&$8.47\%$&$5.58\%$
\\
$L_\text{X}/\text{SFR}$&$10^{40}$ &$8.30\%$ &$9.85\%$
%&$18.10\%$&$13.76\%$
\\ 
\end{tabular}
\end{ruledtabular}
\label{tab:table4}
\end{table}

\section{\label{sec:discussion}degeneracy between fuzzy dark matter effects and astrophysics}

In this section, we examine the degeneracy between the FDM dynamics
and some realistic astrophysical effects.
We evaluate the impact of this degeneracy
on the measurement forecasts of FDM parameters.
For these purposes, we perform \verb|21cmFAST| simulations
based on its mass-dependent model \cite{park2019inferring},
as mentioned in \S\ref{sec:Astro}.
This model adopts a more sophisticated prescription
for cosmic star formation and emission of ionizing photons.

Here we briefly review the mass-dependent model.
The emission rate of ionizing photons now depends on the halo mass as follows:
\begin{IEEEeqnarray}{rCl}
    \dot n_\mathrm{ion} & = &\frac{\ud}{\ud t}
    \bigg[\frac{\zeta_{10}}{\bar\rho_\mathrm{m}} \int_{M_\mathrm{min}}^{\infty} 
    \left(\frac{M}{10^{10}M_\odot}\right)^{\alpha_\mathrm{esc}+\alpha_*}
    \exp\left(-\frac{M_\mathrm{turn}}{M}\right)\quad\nonumber\\
    & & ~~M\frac{\ud n}{\ud M}\,\ud M\bigg],\label{eq:zeta_dep}
\end{IEEEeqnarray}
where $\zeta_{10}\equiv f_\mathrm{esc,10}\,f_\mathrm{*,10}\,N_{\gamma/\mathrm{b}}$
and $N_{\gamma/\mathrm{b}}$ is the number of ionizing photons per stellar baryon.
%Eq.~(\ref{eq:zeta_dep}) 
The above emission rate essentially adopts the following
mass-dependent parametrizations of the escape fraction
and the fraction of galactic gas in stars:
\begin{equation}
\begin{aligned}
    & f_\mathrm{esc}=f_\mathrm{esc,10}
    \left(\frac{M}{10^{10}M_\odot}\right)^{\alpha_\mathrm{esc}},\\
    & f_*=f_{*,10}\left(\frac{M}{10^{10}M_\odot}\right)^{\alpha_*},
\end{aligned}
\end{equation}
where $f_\mathrm{esc,10}$ and $f_\mathrm{*,10}$ are the values
for halos of mass $10^{10}\,M_\odot$.
They are free parameters of the model,
together with the indices $\alpha_\mathrm{esc}$ and $\alpha_*$.
Finally, the factor $\exp(-{M_\mathrm{turn}}/{M})\equiv f_\mathrm{duty}(M)$ in Eq.~(\ref{eq:zeta_dep})
accounts for the quenching of star formation in low-mass halos due to feedbacks,
parametrized by a redshift-independent duty cycle.
%the minimum mass of halos that can produce ionizing photons
%is assumed to be proportional to $M_\mathrm{turn}$, expressed as 
Here $M_\mathrm{turn}$ is also a free parameter of the model.
It further determines the value of ${M}_\mathrm{min}$
by the assumption that ${M}_\mathrm{min}=M_\mathrm{turn}/50$; cf.~\S\ref{sec:Astro}.
Table~\ref{tab:table5} summarizes all the parameters of the mass-dependent model
and their fiducial values.

\begin{table}[ht]%The best place to locate the table environment is directly after its first reference in text
\caption{Free parameters in the mass-dependent model
and their fiducial values for the Fisher matrix forecast.}
\begin{ruledtabular}
\begin{tabular}{ccc}
\textrm{Parameter}&
\textrm{Symbol}&
%\multicolumn{1}{c}{\textrm{value}}&
\textrm{value}\\
\colrule
Normalized FDM Particle Mass &$m_{22}$&10  \\
FDM HMF Index&$\alpha$&-1.1 \\
Gas Fraction Constant &$f_{*,10}$& 0.05\\
Escape Fraction Constant &$f_{\mathrm{esc},10}$& 0.1  \\
Gas Fraction Index &$\alpha_{*}$& 0.5\\
Escape Fraction Index &$\alpha_\mathrm{esc}$& -0.5  \\
Turnover Mass &$M_\text{turn}$&$5\times10^8$ \\
Soft-band x-ray luminosity &$L_\text{X}/\text{SFR}$&$10^{40}$ \\
\end{tabular}
\end{ruledtabular}
\label{tab:table5}
\end{table}

\begin{figure}[ht]
\includegraphics[width=8cm]{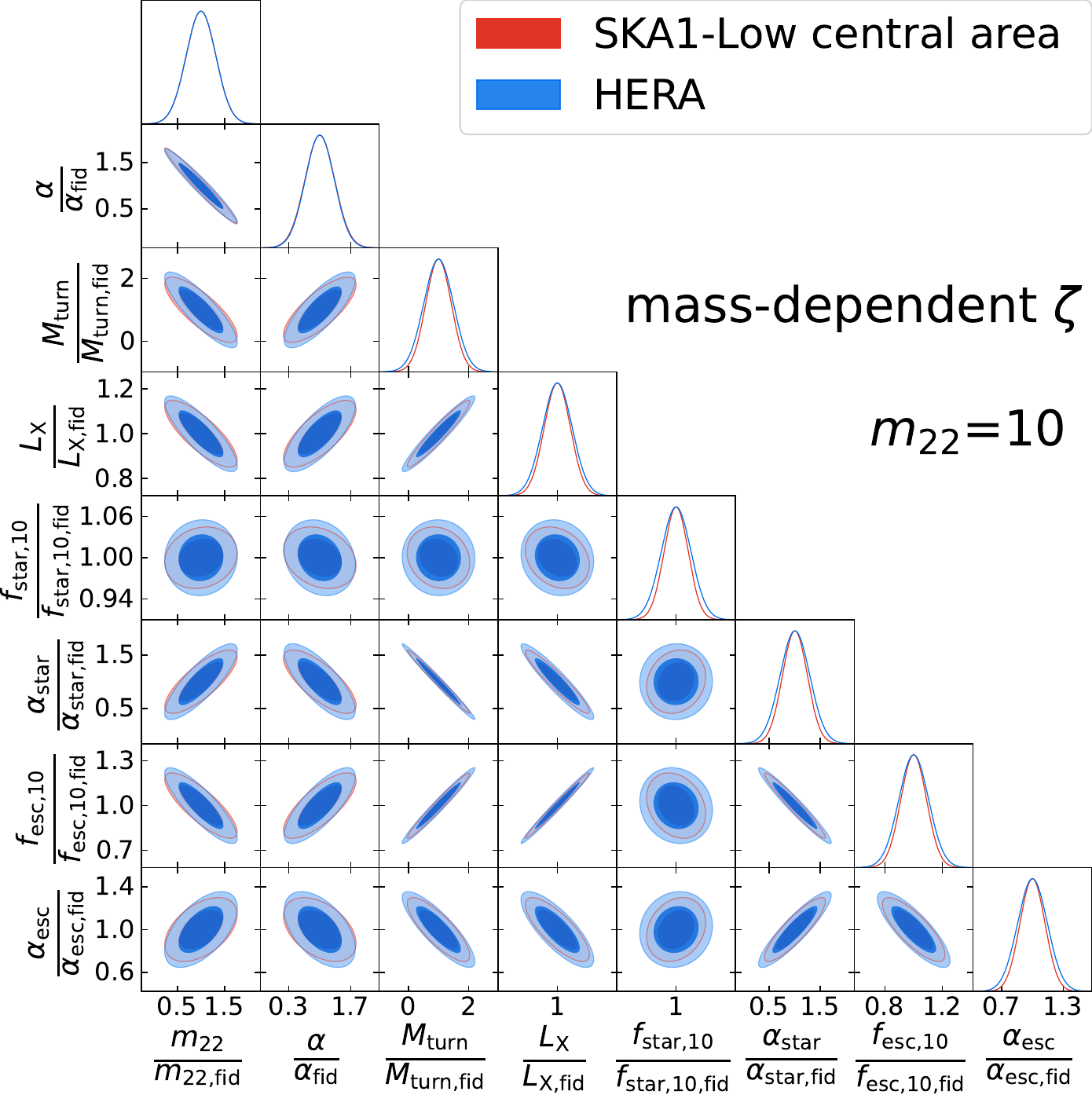}% Here is how to import EPS art
\caption{Similar to Fig.~\ref{fig:nz},
but for the mass-dependent model
in which the ionization efficiency depends on the halo mass.}
\label{fig:z10}
\end{figure}

The results of the Fisher matrix analysis are illustrated in Fig.~\ref{fig:z10}
and the 1$\sigma$ constraints on the model parameters shown in Table~\ref{tab:table6}.
We find that for the mass-dependent model,
the boson mass can only be constrained within 63.0\%/63.2\% of the fiducial value
at 95\% confidence by SKA1-Low central area/HERA,
significantly worse than those for the fiducial mass-independent model.
These results demonstrate the degeneracy between the FDM physics and the CD/EoR astrophysics.

\begin{table}[ht]%The best place to locate the table environment is directly after its first reference in text
\caption{Similar to Table~\ref{tab:table4},
but for the mass-dependent model.
%The parameters constrained by HERA and SKA1-Low central area in FDM universe with $m_\text{22}=10$ which ionization efficiency is mass dependent. The constraint ability we shown in this table is 1$\sigma$ range.
}
\begin{ruledtabular}
\begin{tabular}{cccc}
\textrm{Parameter}&
\textrm{value}&
%\multicolumn{1}{c}{\textrm{value}}&
\textrm{SKA1-Low central area (1$\sigma$)}&
\textrm{HERA(1$\sigma$)} \\
\colrule
$m_{22}$&10& $31.51\%$ &$31.60\%$  \\
$\alpha$&-1.1& $33.61\%$&$32.88\%$\\
$f_{*,10}$&0.05& $1.83\%$& $2.25\%$ \\
$f_{esc,10}$&0.1& $8.95\%$& $10.38\%$ \\
$\alpha_{*}$&0.5& $24.30\%$& $29.07\%$ \\
$\alpha_{esc}$&-0.5& $12.19\%$& $14.48\%$ \\
$M_\text{turn}$&$5\times10^{8}$&$42.42\%$& $49.26\%$ \\
$L_\text{X}/\text{SFR}$&$10^{40}$ &$6.09\%$ & $6.91\%$ \\ 
\end{tabular}
\end{ruledtabular}
\label{tab:table6}
\end{table}

\begin{figure}[ht]
\includegraphics[width=8.5cm]{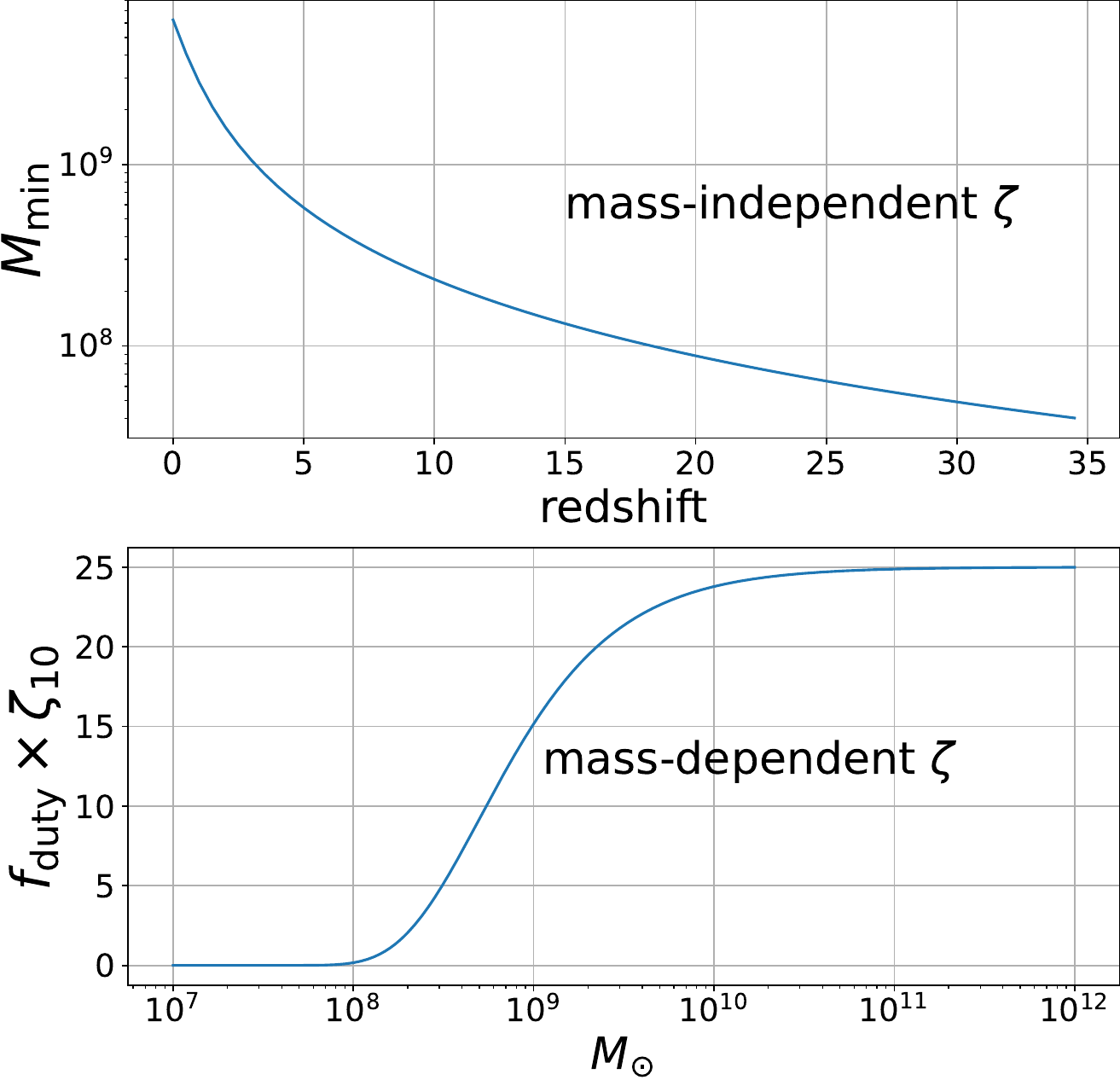}% Here is how to import EPS art
\caption{Degeneracy between the FDM and the astrophysical effects
in the mass-dependent model explained.\\
\emph{Upper}: Redshift evolution of $M_\mathrm{min}$ in the fiducial mass-independent model,
roughly corresponding to the smallest atomic cooling halos.
In this case, $M_\mathrm{min}$ is a subdominant factor compared with the turnover in the FDM HMF.\\
\emph{Lower}: Duty cycle as a function of the halo mass in the fiducial mass-dependent model.
The quenching of star formation due to feedbacks serves an important contribution
to the small-scale depletion of ionizing photons, comparable to the FDM effects.}
\label{fig:model}
\end{figure}

\begin{figure*}[t]
\includegraphics[width=19cm]{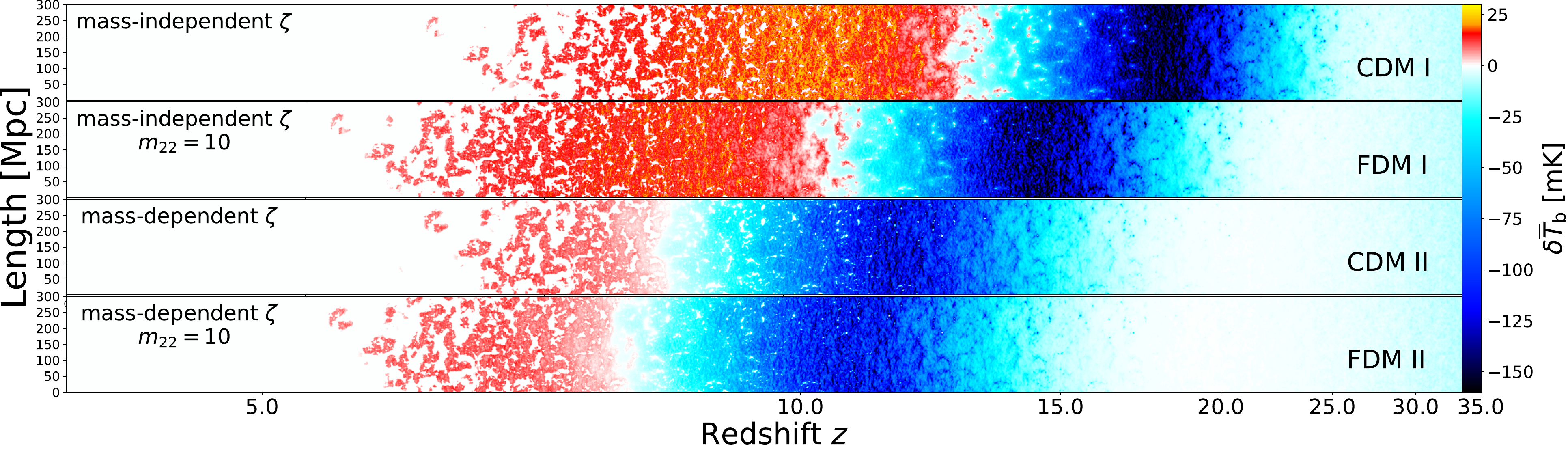}% Here is how to import EPS art
\caption{Light cone slices of the 21-cm brightness temperature during CD/EoR.
The upper two panels illustrate a CDM universe (CDM I)
and an FDM universe (FDM I) in the mass-independent model.
The lower two panels, CDM II and FDM II,
illustrate two universes in the mass-dependent model.
In both CDM I and FDM I, we set $\zeta=25$ so that it matches
the upper limit of $f_\mathrm{duty}(M)\times\zeta_{10}$ in the mass-dependent cases. 
% Are the mass-dependent cases fiducial?
%(see the lower panel in Fig.~\ref{fig:model}).
In both FDM cases, the boson mass is set to $m_\text{22}=10$.}
\label{fig:lcnformodel}
\end{figure*}

Figure~\ref{fig:model} provides more insights into the degeneracy.
The upper panel shows that for the fiducial mass-independent model with $T_\text{vir}=2\times 10^4\,\text{K}$,
$M_\text{min}$ is always smaller than $\sim10^{8.5}\,M_\odot$ throughout CD/EoR.
Since this is the mass scale below which FDM halo formation is suppressed (cf.~Fig.~\ref{fig:HMF}),
the FDM HMF is therefore the dominant factor in the small-scale deficiency of ionization photons,
in contrast to $M_\text{min}$.
As a result, the delaying effect on the 21-cm signals
is mostly attributed to the FDM dynamics.
%\update{so note that the model is not completely mass independent.}
However, the situation changes in the mass-dependent model.
%By observing the upper panel of Fig.~\ref{fig:model},
%we can see that $M_\text{turn}$ is gradually higher than the inhibition of FDM effect in the post-reionization era, that is to say, in our fiducial model, the suppression of FDM effect on the formation of dark matter halo is very effective throughout CD/EoR. 
%Let's look at the lower panel of Fig.~\ref{fig:model}, which represents a mass-dependent model of ionization efficiency.
In its fiducial model, $\alpha_*+\alpha_\mathrm{esc}=0$
so that the \emph{astrophysical} mass dependence of the emission rate of ionizing photons in Eq.~(\ref{eq:zeta_dep})
is entirely sourced by the term $f_\text{duty}(M)\times \zeta_{10}$,
which is illustrated in the lower panel of Fig.~\ref{fig:model}.
It shows that the duty cycle is exponentially suppressed
below mass scales $\sim10^{8.5}\,M_\odot$.
Thus, small-scale quenching of star formation
becomes a competitive factor in reducing the number of ionizing photons,
along with the FDM HMF.
This explains why the FDM effects are degenerate with astrophysical effects
in the more realistic mass-dependent model,
as they both contribute to the delaying effect.
Hence, the 21-cm power spectrum becomes
less sensitive to the FDM parameters in this model.

To visualize the impact of the mass-dependent model,
we present a full comparison between the 21-cm brightness temperature light cones in different cases,
as shown in Fig.~\ref{fig:lcnformodel}.
%our fiducial model and mass-dependent $\zeta$ model under CDM and FDM universes, with $\zeta$ set to 25 in the fiducial model to match the formal $\zeta$ of the mass-dependent model, 
In the two cases with mass-independent $\zeta$,
the FDM model (FDM I) demonstrates a significant delay in the 21-cm signal
relative to the CDM model (CDM I).
By contrast, the two cases in the mass-dependent model, FDM II and CDM II,
exhibit less pronounced differences due to the degeneracy.
On the other hand, cosmic reionization in CDM II
is considerably delayed compared with CDM I
solely because of the astrophysical effects,
which appear to be even stronger than the effects of FDM dynamics in FDM I
with $m_{22}=10$.
%Additionally, we can find in this model, FDM extends the epoch of reionzation in space scale, unlike what we see in the fiducial model.

\section{\label{sec:conclusion}CONCLUSION}

In this work, we explored the impacts of fuzzy dark matter
on the 21-cm signals from cosmic dawn and the epoch of reionization.
We considered full FDM dynamics in both linear and nonlinear regimes.
The FDM linear matter power spectrum and the (nonlinear) FDM halo mass function
are implemented in the seminumerical simulation code \verb|21cmFAST|.
In particular, we adopted the fitting formula for the global FDM HMF
from full numerical simulations
and devised a novel ansatz to model variations of the FDM HMF
in different density environments.
The simulated light cones of the 21-cm brightness temperature during CD/EoR
are thus determined by the formation of first luminous objects
and the reionization process in various FDM cosmologies.
We performed Fisher matrix forecasts
to examine the prospects of constraining the FDM parameters, $(m_{22},\,\alpha)$,
using the ongoing/upcoming 21-cm power spectrum measurements,
HERA and SKA1-Low (central area).
Our main results are as follows:
\setdefaultleftmargin{0.cm}{}{}{}{}{}
\begin{enumerate}

\item We presented a comprehensive analysis
of the delaying effect on the 21-cm signals due to the FDM dynamics
from multiple perspectives.
In fact, all three signature epochs during CD/EoR
(Ly$\alpha$ coupling, X-ray heating, reionization)
are not only delayed but also shortened,
as shown by the rises and falls in the evolutions
of the global 21-cm signal and the 21-cm power.
The shortening effect is particularly reflected in the diminishing
of the early trough in the $\Delta^2_{21}(k)$ evolution for small boson masses,
which roughly marks the transition
between the Ly$\alpha$ coupling epoch and the x-ray heating epoch.

\item We investigated the influence of taking into account the nonlinear FDM dynamics
by comparing the full FDM model with the intermediate ``FDM I.C.s'' model.
The differences in the $\Delta^2_{21}(k)$ evolution
are only significant when x-ray heating begins.
It can be implied that FDM effects on CD/EoR
are dominated by its linear dynamics at early times
while the nonlinear wave dynamics become important at late times.

%We present the first Fisher matrix forecast of the 21-cm power spectrum generated by a FDM model with HMF based on numerical simulation varying FDM cosmology and astrophysical parameters. 
\item We find that the constraints on the FDM parameters
from SKA1-Low (central area) and HERA are comparable.
Assuming a mass-independent ionizing efficiency, $\zeta$,
and a moderate treatment for the foreground wedge,
the 21-cm power spectrum measurements
from both experiments should be able to determine the boson mass
to within $\sim10$\% at $2\sigma$ confidence
in the fiducial FDM universe with $m_\text{22}=10$.

\item More realistic modeling of the astrophysical sector
with a mass-dependent ionizing efficiency
results in a degeneracy between the FDM physics and the astrophysics.
Taking account of this degeneracy, we find that
the predicted constraints become about six times worse.
Therefore, it will be challenging to test the FDM cosmology
using 21-cm power spectrum data alone in realistic settings.
Combined analyses including the global 21-cm signal or future 21-cm tomographic images
will be necessary for probing the FDM (or other alternative dark matter) scenario.
%Additionally, we find that in this model, the FDM effect extends the reionization epoch on spatial scales, which differs from what we see in the fiducial model.

\end{enumerate}

\section*{ACKNOWLEDGMENTS}

Y.M. is supported by the National SKA Program of China (Grant No.~2020SKA0110401).
B.L. is supported by the Guangxi Natural Science Foundation (Grant No.~2023GXNSFBA026114)
and the National Natural Science Foundation of China (Grant Nos.~12203012 and 12494575).
Additional support was provided by the Guangxi Talent Program (“Highland of Innovation Talents”).

%\appendix

%\section{Supplement}
%We use the first-order Zel'dovich approximation and the redshift space distortion (RSD) effect when generating the density field with 21cmFAST, noting that this is not accurate enough for an FDM dominated universe. The mass-dependent model we discussed in the previous section does not include simulations of the velocity acoustic oscillations (VAO) effect and Lyman-Werner (LW) effect from first galaxies \cite{munoz2022impact,qin2020tale}, because the current VAO effect does not apply to the FDM universe.

%\nocite{*}

\section*{DATA AVAILABILITY}

The data that support the findings of this article are not publicly available upon publication because it is not technically feasible and/or the cost of preparing, depositing, and hosting the data would be prohibitive within the terms of this research project. The data are available from the authors upon reasonable request.

\bibliography{21cm}% Produces the bibliography via BibTeX.

\end{document}